\documentstyle[10pt,emulateapj]{article}

\input{epsf}

\journalid{}{}
\articleid{}{}
\begin{document}
\def\lax    {\ifmmode{_<\atop^{\sim}}\else{${_<\atop^{\sim}}$}\fi}
\def\gax    {\ifmmode{_>\atop^{\sim}}\else{${_>\atop^{\sim}}$}\fi}
\def\gtorder{\mathrel{\raise.3ex\hbox{$>$}\mkern-14mu
             \lower0.6ex\hbox{$\sim$}}}
\def\ltorder{\mathrel{\raise.3ex\hbox{$<$}\mkern-14mu
             \lower0.6ex\hbox{$\sim$}}}
 
\long\def\***#1{{\sc #1}}
 
\title{M31 Globular Cluster X-ray sources: {\em XMM-Newton} and {\em Chandra} results.}

\author{Sergey Trudolyubov\altaffilmark{1,2,3} and William Priedhorsky\altaffilmark{2}}

\altaffiltext{1}{IGPP, University of California, Riverside, CA 92521}

\altaffiltext{2}{Los Alamos National Laboratory, Los Alamos, NM 87545}

\altaffiltext{3}{Space Research Institute, Russian Academy of Sciences, 
Profsoyuznaya 84/32, Moscow, 117810 Russia}

\begin{abstract}

We present the results of M31 globular cluster (GC) X-ray source survey, based on the data of 
{\em XMM-Newton} and {\em Chandra} observations covering $\sim 6100$ arcmin$^{2}$ of M31. We 
detected 43 X-ray sources coincident with globular cluster candidates from various optical 
surveys. 

The inferred isotropic X-ray luminosities of GC sources lie between $\sim 10^{35}$ and 
$\sim 10^{39}$ erg s$^{-1}$ in the $0.3 - 10$ keV energy band. The spectral properties of 31 
brightest sources from our sample were found to be similar to that of the low mass X-ray binaries 
located in the bulge and globular clusters of the Milky Way Galaxy. The spectral distribution of 
M31 GC X-ray sources is consistent with that derived for the bulge of M31 and other nearby galaxies 
of different morphological type. Several sources demonstrate a correlation between the level of 
X-ray flux and hardness of their energy spectrum reminiscent of the Galactic Z and atoll sources. 

We found that $\sim 80\%$ of the M31 GC sources with multiple flux measurements available show 
significant variability on a time scales from days to years. The X-ray source RX J0043.2+4127, 
coincident with GC Bo 163, has been found to show recurrent transient outbursts with peak 
luminosities of $\sim 10^{38}$ ergs s$^{-1}$. Several sources in our sample show significant 
variability on a time scale of individual observations, ranging from aperiodic fluctuations to 
regular dipping.

The X-ray luminosity function of GC sources is found to be significantly different from that of 
the point sources in the bulge and disk of M31. The luminosity distribution of M31 GC sources has 
$\sim 10$ times higher peak luminosity and much higher fraction of bright sources than the Milky 
Way GC distribution. Six persistent sources in our sample (or $\sim 14 \%$ of the total number) 
have luminosities exceeding $10^{38}$ ergs s$^{-1}$ during all observations, and three other sources 
occasionally exceed that luminosity level. Our observations indicate that GC sources make dominant 
contribution to the bright source counts in the areas of M31 covered by the survey: $\sim 40 \%$ 
of the total number of sources with luminosities above $10^{37}$ ergs s$^{-1}$ reside in GCs with 
fraction of GC sources rising to $67 - 90\%$ for the luminosities above $10^{38}$ ergs s$^{-1}$. 
The contribution of the GC sources to the total number of bright sources found in M31 is much higher 
than in the Milky Way galaxy, but surprisingly close to the early-type galaxies. We found that 
brightest M31 GC sources tend to reside at large galactocentric distances outside the central bulge.

We found that globular clusters hosting bright X-ray sources are optically brighter and more metal 
rich than the rest of M31 globular clusters, in agreement with previous studies. The brightest 
sources with luminosities above $\sim 10^{38}$ ergs s$^{-1}$ show tendency to reside in more metal 
poor clusters.

The remarkable similarities between the properties of the M31 GC X-ray sources and that of the Galactic 
neutron star LMXBs allow us expect most of the persistent M31 GC X-ray sources to be LMXB systems with 
neutron star primaries. However, the current X-ray spectral and timing data can not rule out the 
possibility of finding an active accreting black holes in our GC source sample.

\end{abstract} 

\section{INTRODUCTION}
Globular clusters (GCs) provide us with crucial information on galaxy structure and formation 
mechanisms. They are also ideal laboratories for studying stellar populations and evolution of 
dense stellar systems. X-ray surveys revealed a number of bright X-ray sources associated with 
Milky Way (MW) GCs and identified with low-mass X-ray binaries (LMXBs) (\cite{HG83,V95}). The 
ratio of LMXBs to stellar mass is two orders of magnitude higher for GCs than for the rest of 
our Galaxy (\cite{Liu01}). This overabundance of LMXB has been explained by dynamical effects 
such as tidal capture and three/four body interactions in the high stellar density environment 
(\cite{Clark75,FPR75}). The number of known Galactic GCs hosting bright X-ray sources 
($L_{X} > 10^{36}$ ergs s$^{-1}$) is pretty small: there are fourteen such systems (\cite{Liu01}). 
The presence of Type I X-ray bursts indicates that all luminous GC sources contain accreting neutron 
stars rather than black holes. With the advent of the modern X-ray observatories, it has become 
possible to study the X-ray properties of GCs associated with the nearby galaxies (\cite{Angelini01}; 
\cite{DiStefano01}; \cite{Kundu02}). Based on a large number of objects, the extragalactic GC 
X-ray sources allow a broad range of statistical population studies, impossible with a limited 
sample of Galactic GC X-ray sources.

The Andromeda Galaxy (M31), the closest giant spiral galaxy to our own, is a unique object for 
the study of optical and X-ray astronomy. Its proximity and favorable orientation allow to observe 
stellar populations over the full extent of the galaxy at a nearly uniform distance, and with less 
severe effects of line-of-sight contamination from interstellar gas and dust. Due to similarities 
between the two galaxies, the results from the study of M31 provide an important benchmark for 
comparison with the results from the study of our own Milky Way Galaxy. The Andromeda Galaxy possesses 
a significantly more populous globular cluster system than the Milky Way with more than 430 confirmed 
candidate members (\cite{Barmby00}). Optical studies of the Milky Way and M31 globular cluster 
systems have revealed that these two populations exhibit some remarkable similarities (\cite{vdB00}). 

M31 was observed extensively with {\em Einstein}, {\em ROSAT}, {\em Chandra} and {\em XMM} missions, 
detected hundreds of sources, with bright GC X-ray sources among them (\cite{TF91}; \cite{Primini93}; 
\cite{Supper97}; \cite{Shirey01}). Twenty-one X-ray sources discovered with {\em Einstein} were 
tentatively identified with GCs. Based on these early results, it has been noted that the luminosities 
of the M31 GC sources are higher than any observed for Galactic GCs (\cite{LvS83}), and that the 
fraction of M31 GCs with X-ray sources might be larger than the fraction in the Milky Way (\cite{Bo87}). 
Observations of the central bulge of M31 with {\em ROSAT}/HRI detected 18 GC X-ray sources and placed 
upper limits on the X-ray flux from 32 other globular clusters (\cite{Primini93}). Primini, Forman 
\& Jones (1993) found that the luminosity distribution of M31 GC sources contains a larger population 
of high luminosity sources than that of the Milky Way GC X-ray sources, although the peak luminosities 
of the two distributions were roughly comparable. The {\em ROSAT}/PSPC survey covering most of M31 
revealed 31 X-ray sources coincident with M31 GC candidates (Supper et al. 1997, 2001). Based on this 
larger sample, Supper et al. (1997, 2001) concluded that there is no significant difference between the 
luminosity distributions of the M31 GC X-ray sources and their Galactic counterparts. Recent {\em Chandra} 
observations covering central bulge of and two regions in the northern and southern disk revealed 28 GC 
X-ray sources with 15 of them newly discovered (\cite{DiStefano01}). Approximately $\sim 30\%$ of all GC 
sources detected in {\em Chandra} observations have $0.5 - 7$ keV luminosities above $10^{37}$ ergs 
s$^{-1}$ and $\sim 10\%$ of all the sources have luminosities of $10^{38}$ ergs s$^{-1}$ and higher. 
Both the peak X-ray luminosity and relative fraction of the bright sources with $L_{X} > 10^{37}$ ergs 
s$^{-1}$ were found to be significantly higher for the {\em Chandra} GC source sample than for the GC 
sources in our Galaxy (\cite{DiStefano01}) in contrast to the earlier {\em ROSAT} results 
(Supper et al. 1997, 2001). The limited sensitivity of {\em Chandra} observations allowed to study 
spectral properties and variability of only five most luminous sources found to be consistent with 
properties of the Galactic sources. 

We present the results of M31 globular cluster (GC) X-ray source survey, based on the data of 
{\em XMM-Newton} and {\em Chandra} observations covering $\sim 6100$ arcmin$^{2}$ of the Andromeda 
galaxy (M31). The unique combination of the unprecedented sensitivity of {\em XMM-Newton} and superior 
spatial resolution of {\em Chandra} observations allows accurate localization and detailed study of 
spectral and timing properties of the GC X-ray sources in both central regions of M31 and at large 
galactocentric distances. We concentrate on the analysis of spectral properties and variability of 
individual bright sources and study a group properties of all GC X-ray sources detected in our survey. 
We study the relation between the properties of the X-ray sources and that of the globular clusters 
hosting them. Finally, the properties of M31 GC X-ray sources are compared to the properties of 
low-mass X-ray binaries LMXB in M31, Milky Way and other galaxies of different morphological types.     

\section{OBSERVATIONS AND DATA ANALYSIS}

\begin{figure}
\epsfxsize=18cm
\epsffile{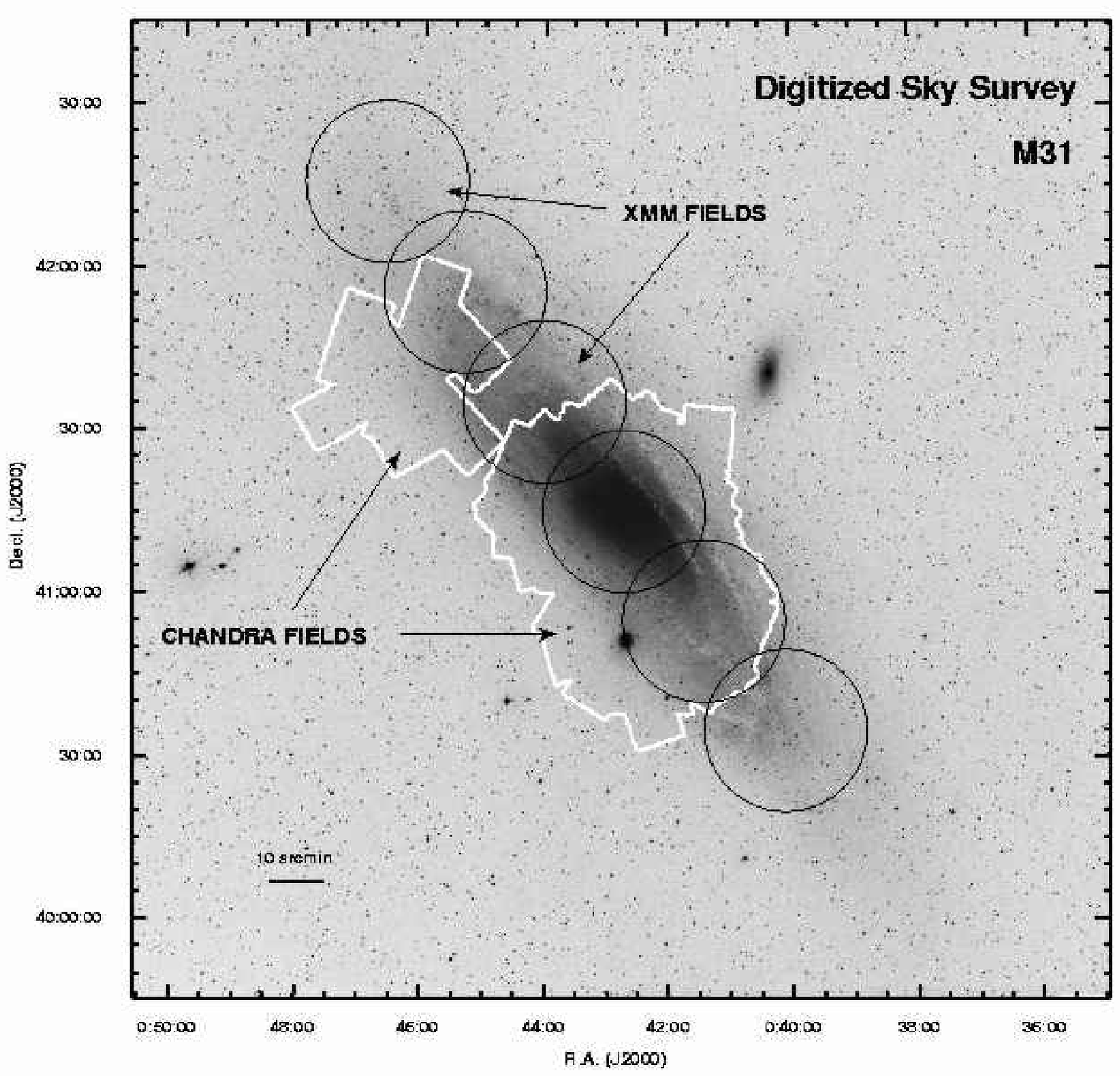}
\caption{Optical image of M31 from the Digitized Sky Survey with regions covered by 
{\em XMM-Newton}/EPIC ({\em black circles}) and {\em Chandra}/ACIS/HRC observations 
({\em white regions}). The remote globular cluster G1 field is located outside image 
boundary. 
\label{image_opt_xray}}
\end{figure}

In the following analysis we used the data of 10 {\em XMM-Newton} observations of various 
M31 fields (Table \ref{obslog_xmm}; Fig. \ref{image_opt_xray}). All {\em XMM-Newton} observations 
described in this paper were performed as a part of the Performance Verification and Guaranteed 
Time Programs (\cite{Shirey01,lum_distr02,dipper02}). Four {\em XMM} observations cover the central 
bulge of M31, five -- its northern and southern disk regions, and one -- a distant M31 globular 
cluster G1 (Mayall II) field. In the following analysis we use the data of three European Photon 
Imaging Camera (EPIC) instruments: two EPIC MOS detectors (\cite{Turner01}) and EPIC-pn detector 
(\cite{Strueder01}). During all observations EPIC instruments were operated in the {\em full window} 
mode ($30\arcmin$ FOV) with {\em medium} or {\em thin} optical blocking filter. 

We reduced EPIC data with the {\em XMM-Newton} Science Analysis System (SAS v 5.3)
\footnote{See http://xmm.vilspa.esa.es/user}. We performed standard screening of the 
EPIC data to exclude time intervals with high background levels. Images in the celestial coordinates 
with a pixel size of 2$\arcsec$ have been accumulated in the $0.3 - 7.0$ keV energy band for the 
EPIC-MOS1, MOS2 and pn detectors. To generate lightcurves and spectra of X-ray sources, we used 
elliptical extraction regions with semi-axes size of $\sim 20 - 80 \arcsec$ (depending on the 
distance of the source from the telescope axis) and subtracted as background the spectrum of 
adjacent source-free regions with subsequent normalization by a ratio of the detector areas. We 
corrected the count rates of the sources for the vignetting of the XMM telescope, based on the 
Current Calibration Files.We used data in the $0.3 - 10$ keV energy band because of the uncertainties 
in the calibration of the EPIC instruments outside this range. All fluxes and luminosities derived 
from spectral analysis apply to this band, unless specified otherwise. We used spectral response 
matrices generated by XMM SAS tasks. The EPIC count rates were converted into energy fluxes in 
the $0.3 - 10$ keV energy band using analytical fits to the spectra for brighter sources or  
PIMMS\footnote{See http://heasarc.gsfc.nasa.gov/Tools/w3pimms.html}, assuming standard parameters: 
an absorbed simple power law model with $N_{\rm H} = 7 \times 10^{20}$ cm$^{-2}$ and photon index 
$\alpha = 1.7$ (\cite{Shirey01}).

We analyzed a series of publicly available {\em Chandra} data for M31 fields containing optically 
identified GC candidates consisting of 36 observations with the Advanced CCD Imaging Spectrometer 
(ACIS) and 3 High Resolution Camera (HRC) observations (Table \ref{obslog_chandra}; Fig. 
\ref{image_opt_xray}). A more detailed description of these observations can be found in Garcia et al. 
(2000), DiStefano et al. (2002), Kong et al. (2002) and Kaaret et al. (2002). 

The data of {\em Chandra} observations was processed using the 
CIAO v3.0\footnote{http://asc.harvard.edu/ciao/} threads. We performed standard screening 
of the {\em Chandra} data to exclude time intervals with high background levels. Images in the 
celestial coordinates with a pixel size of 1$\arcsec$ have been accumulated in the $0.3 - 7.0$ 
keV energy band for source detection procedure. Only data in the $0.3 - 7.0$ and $0.5 - 7.0$ keV 
energy ranges was used in the spectral analysis of ACIS-S and ACIS-I observations. To estimate 
energy fluxes and spectra, we extracted counts within source elliptical region with semi-axes 
of $3\arcsec - 40\arcsec$ (depending on the source distance from the telescope axis). Background 
counts were extracted from the adjacent source-free regions with subsequent normalization by ratio 
of detector areas. To account for the continuous degradation in the effective low-energy quantum 
efficiency of the ACIS detectors\footnote{http://cxc.harvard.edu/cal/Acis/Cal prods/qeDeg/}, 
we performed correction of the spectral response files applying {\em ACISABS} model to the ARF 
files. For a few brightest GC sources pileup fraction in their ACIS spectra is estimated to be at 
the level of $\sim 10 - 30\%$. In some cases, to correct for a pileup effect in the ACIS spectra, 
we extracted source spectrum from the PSF wings, where the source count rate is low enough that 
pileup effect can be ignored. We also used Sherpa\footnote{http://cxc.harvard.edu/sherpa/} 
implementation of the pileup model by Davis (2001,2003). The ACIS count rates were converted into 
energy fluxes in the $0.3 - 10$ keV energy band using analytical fits to the spectra for brighter 
sources or using PIMMS with spectral parameters derived from more sensitive observations or standard 
parameters otherwise. For the HRC count rates the source fluxes were estimated with PIMMS, based on 
the observed count rates and spectral parameters measured with {\em XMM}/EPIC and {\em Chandra}/ACIS.

For both {\em XMM-Newton} and {\em Chandra} data, X-ray sources were detected with the program 
based on the wavelet decomposition algorithm, set at a $4\sigma$ threshold. For our current 
analysis of {\em XMM-Newton} observations, we expect error in the source position determination 
to be dominated by residual systematic error of the order $1 - 3 \arcsec$.

We performed a spectral and timing study of the brightest X-ray sources with total number 
of counts larger than 300. Spectra were grouped to contain a minimum of 20 counts per 
spectral bin and fit to analytical models using the XSPEC 
v.11\footnote{http://heasarc.gsfc.nasa.gov/docs/xanadu/xspec/index.html} fitting package 
(\cite{arnaud96}). EPIC-pn, MOS1 and MOS2 data were fitted simultaneously, but with 
independent normalizations.

We studied timing properties of all bright X-ray sources detected in our observations. 
Fourier power density spectra (PDS) were produced using the lightcurves in the 
$0.3 - 7.0$ keV energy band. Then we performed folding analysis in the vicinity of the 
frequency peaks identified from PDS. We used standard XANADU/XRONOS v.5
\footnote{http://heasarc.gsfc.nasa.gov/docs/xanadu/xronos/xronos.html} tasks to perform 
analysis of the timing properties of bright X-ray sources.

In the following analysis we assume a source distance of 760 kpc (van den Bergh 2000).

\section{M31 GC X-ray sources: identification.}
The positions of 43 X-ray sources detected in the {\em XMM-Newton} and {\em Chandra} 
observations of M31 are consistent with globular cluster candidates from optical surveys. We 
used the following catalogs for identifying GC X-ray source candidates: the Bologna catalog 
(\cite{Bo87}), the catalog by Magnier (1993), and the HST globular cluster candidate catalog 
(\cite{Barmby01}) -- with search radius of $3\arcsec$. The expect that $\sim 4$ of our GC 
X-ray candidates could be spurious matches. The information on the properties of the X-ray sources 
and their optical identifications is shown in Table \ref{source_ID}. 

\begin{figure}
\epsfxsize=18cm
\epsffile{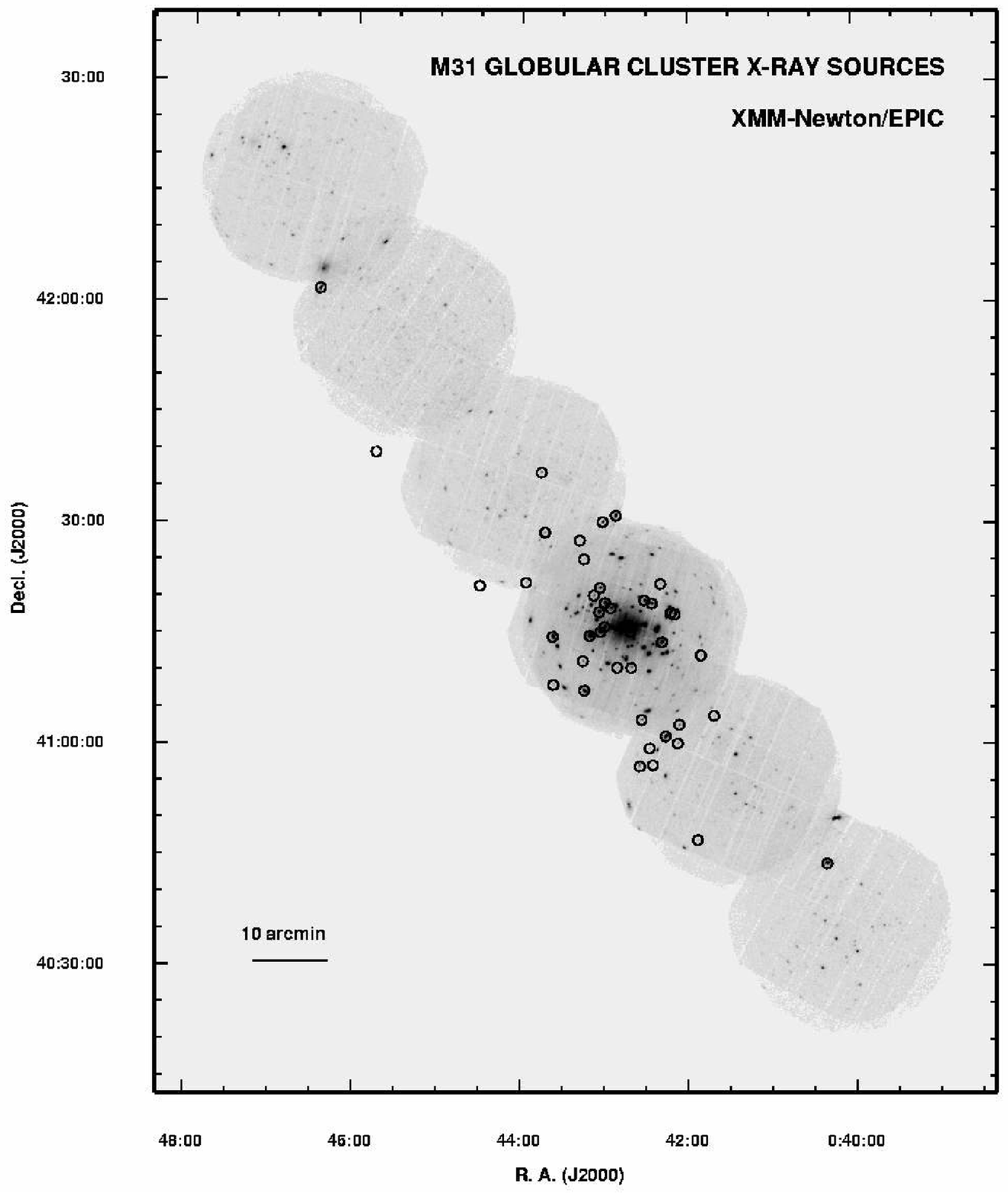}
\caption{Combined {\em XMM}/EPIC images of M31 fields in the $0.3 - 7.0$ keV energy 
band. The image was convolved with circularly symmetric Gaussian function with 
$\sigma = 8 \arcsec$. The globular cluster X-ray sources detected in {\em XMM-Newton} 
and {\em Chandra} observations are marked with circles. 
\label{image_xmm}}
\end{figure}

\begin{figure}
\epsfxsize=18cm
\epsffile{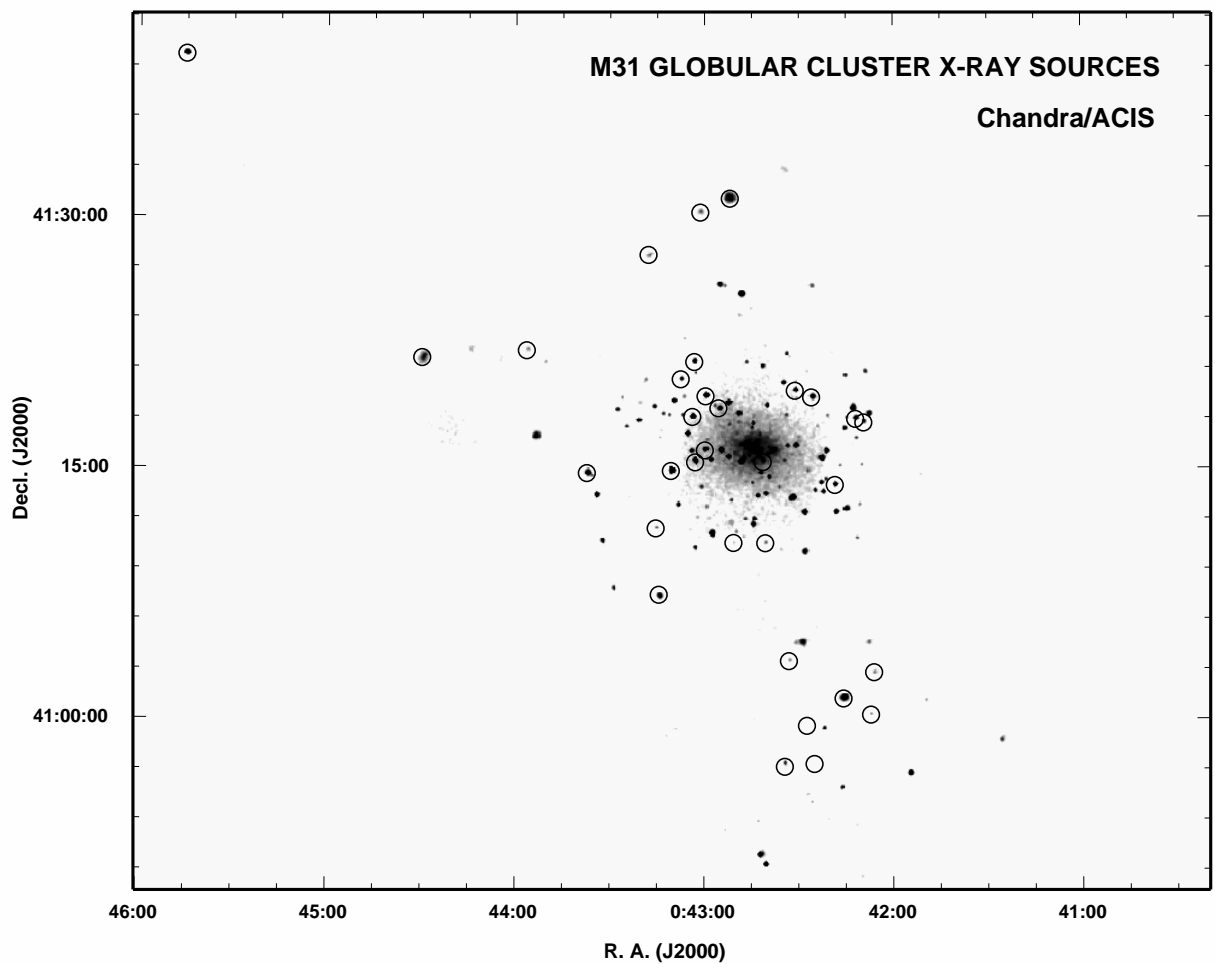}
\caption{Combined {\em Chandra}/ACIS images of M31 fields in the $0.3 - 7.0$ keV 
energy band. The image was convolved with circularly symmetric Gaussian function with 
$\sigma = 8 \arcsec$. The globular cluster X-ray sources detected with {\em Chandra} 
are marked with circles. 
\label{image_chandra}}
\end{figure}

The X-ray images of M31 with GC X-ray source positions circled are shown in Fig. \ref{image_xmm} 
and \ref{image_chandra}. The X-ray image shown in Fig. \ref{image_xmm} was constructed combining 
the data of the {\em XMM-Newton}/EPIC-MOS1, MOS2 and pn cameras in the $0.3 - 7.0$ keV energy 
band. The X-ray image shown in Fig. \ref{image_chandra} combines the data of {\em Chandra}/ACIS 
and HRC detectors. 

\section{Note on the effect of multiple unresolved X-ray sources}
Recent {\em Chandra} observation of the Galactic globular cluster M15 has shown two distinct 
luminous X-ray sources separated by less than $3\arcsec$ (\cite{WA01}). The spatial resolution 
of both {\em Chandra} ($\sim 0.5\arcsec$) and {\em XMM} ($\sim 10\arcsec$) is not sufficient to 
resolve most of possible multiple X-ray sources within one cluster in M31. Therefore, some of the 
M31 GC X-ray sources from our sample can be unresolved composites of separate sources 
(\cite{Angelini01}; \cite{DiStefano01}). It should thus be noted, that the spectral blending and 
superposition of variability of individual sources may complicate direct comparisons with Galactic 
globular cluster sources.

\section{Spectral properties of the M31 globular cluster X-ray sources}
\begin{figure}
\vbox{
\begin{minipage}{18.0cm}
\vbox{
\begin{minipage}{18.0cm}
\epsfxsize=18cm
\epsffile{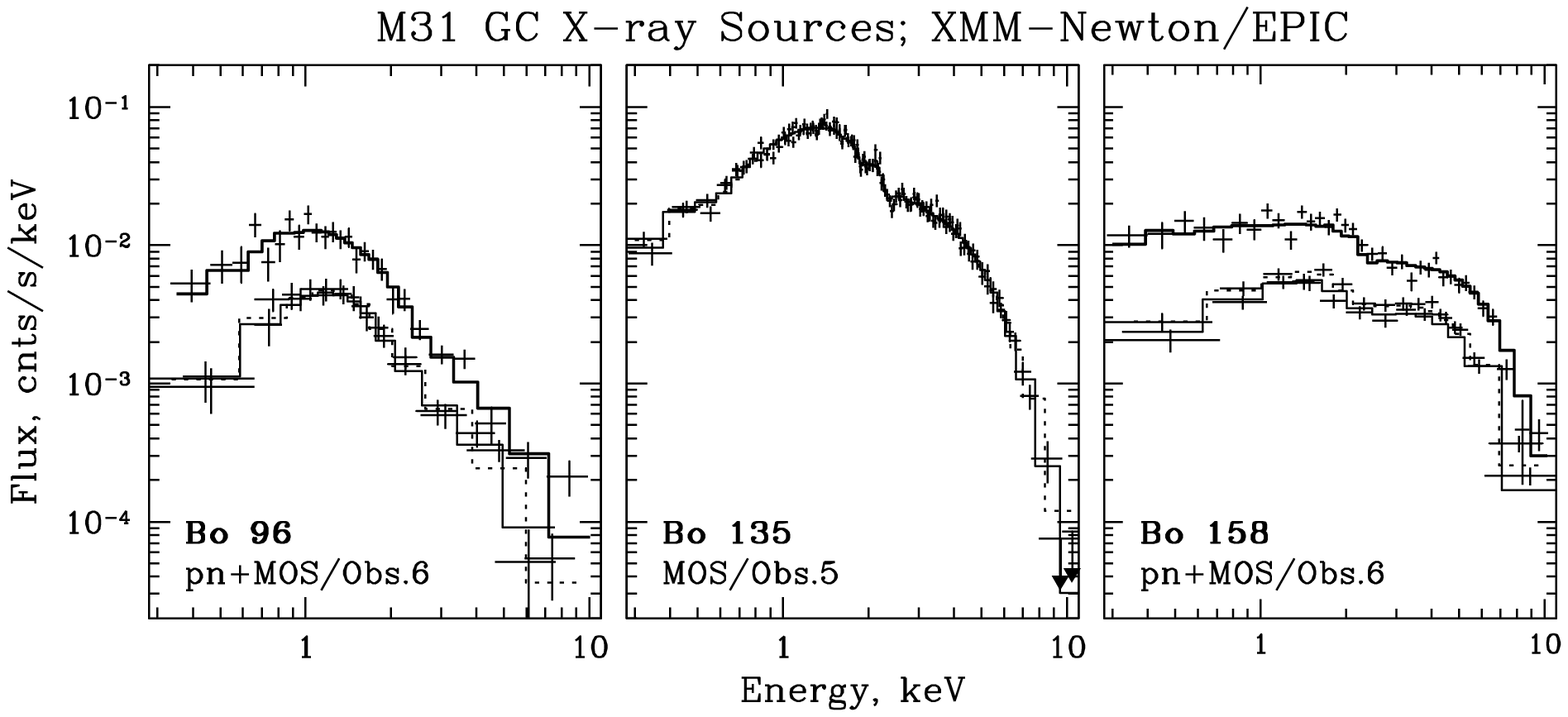}
\end{minipage}
\begin{minipage}{18.0cm}
\epsfxsize=18cm
\epsffile{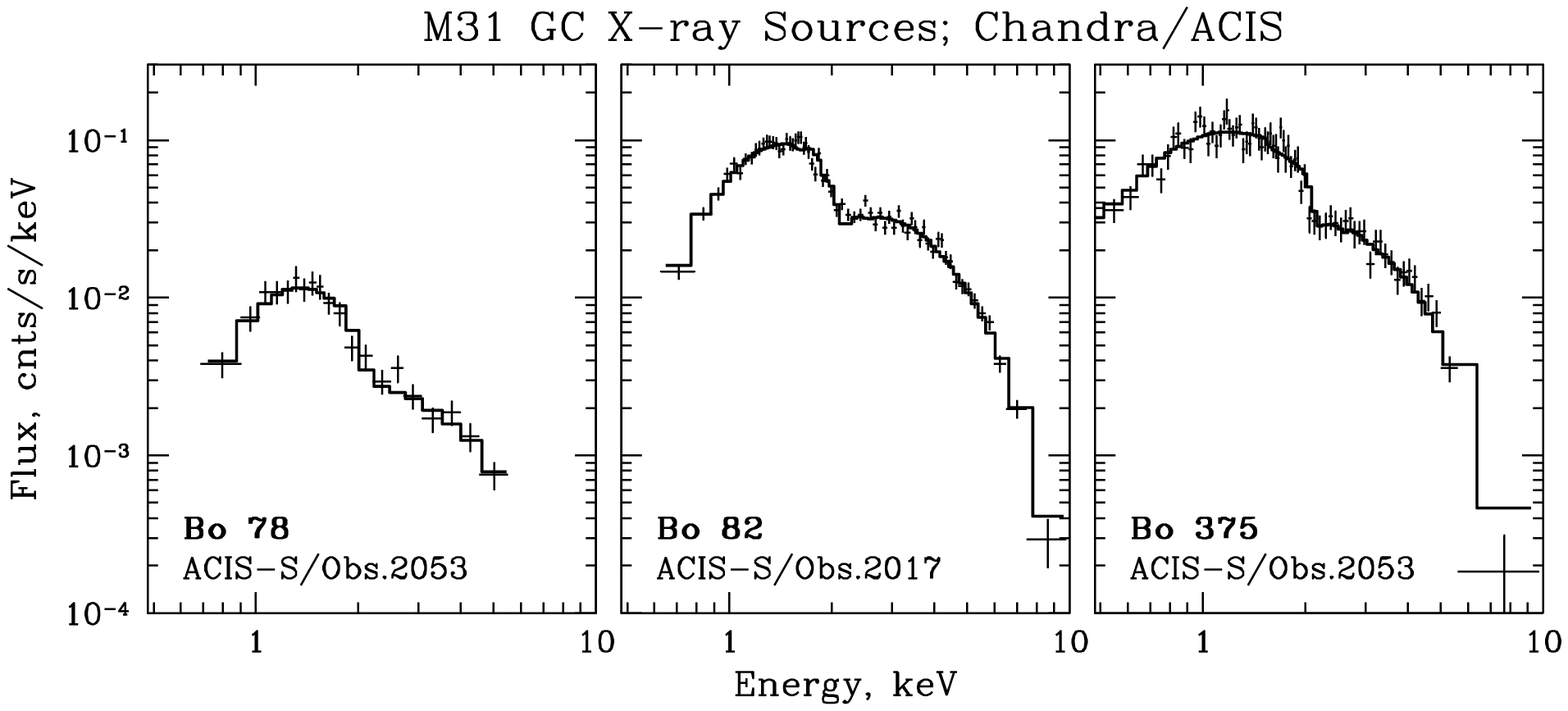}
\end{minipage}
}
\end{minipage}
\begin{minipage}{18.0cm}
\caption{{\em Upper panels}: Representative {\em XMM-Newton}/EPIC spectra of the bright 
GC X-ray sources in the $0.3 - 10$ keV energy band. (EPIC-pn is always the upper spectrum 
and the two MOS spectra overlap). Corresponding model fits to the EPIC-pn, MOS1 and MOS2 
spectra (Tables \ref{spec_par_GCS_xmm_powerlaw}, \ref{spec_par_GCS_cutoffpl_comptt}, 
\ref{spec_par_GCS_two_comp}) are shown with thick, thin and dotted histograms respectively. 
{\em Left panel}: Bo 96 ($\# 14$), EPIC-pn and MOS1,2 data, fit by absorbed simple power law 
model. {\em Middle panel}: Bo 135 ($\# 10$), EPIC-MOS1 and MOS2 data, fit by absorbed power 
law model with exponential cutoff. {\em Right panel}: Bo 158 ($\# 32$), EPIC-pn and MOS1,2 
data, fit by absorbed Comptonization model. {\em Lower panels}: {\em Chandra}/ACIS spectra 
of the bright GC X-ray sources. Model fits are shown with thick histograms (Tables 
\ref{spec_par_GCS_chandra_powerlaw}, \ref{spec_par_GCS_cutoffpl_comptt}, 
\ref{spec_par_GCS_two_comp}). {\em Left panel}: Bo 78 ($\# 9$), ACIS-S data, $0.5 - 7$ keV 
energy range, fit by an absorbed simple power law model. {\em Middle panel}: Bo 82 ($\# 10$), 
ACIS-S data, $0.5 - 10$ keV energy range, fit by an absorbed simple power law model. 
{\em Right panel}: Bo 375 ($\# 42$), ACIS-S data, $0.5 - 10$ keV energy range, fit by an 
absorbed power law model with exponential cutoff. \label{spec_GCS_fig}}
\end{minipage}
}
\end{figure}

Twenty seven and twenty three sources detected with {\em XMM-Newton} and {\em Chandra} have a
sufficient number of counts to allow crude or detailed spectral modeling. Combining the above 
two samples gives us a high-quality information on the spectral properties of 31 bright GC X-ray 
sources. The representative spectra of bright GC sources obtained with {\em XMM-Newton}/EPIC and 
{\em Chandra}/ACIS are shown in Fig. \ref{spec_GCS_fig}.  

The spectra of globular cluster candidates were fitted with a variety of spectral models 
using XSPEC v11. We first considered simple one-component spectral model: an absorbed 
simple power law. The results of fitting this model to the source spectra are given in Tables 
\ref{spec_par_GCS_xmm_powerlaw} and \ref{spec_par_GCS_chandra_powerlaw}. The spectra of all these 
objects can be generally described by an absorbed simple power law model with photon index of 
$\sim 0.8 - 2.8$ and an equivalent absorbing column of $\sim (0.4 - 7.5) \times 10^{21}$ cm$^{-2}$ 
(Table \ref{spec_par_GCS_xmm_powerlaw},\ref{spec_par_GCS_chandra_powerlaw}). The corresponding 
isotropic luminosities of the globular cluster sources differ by four orders of magnitude and fall 
between $\sim 10^{35}$ and $\sim 10^{39}$ ergs s$^{-1}$ in the $0.3 - 10$ keV energy band, assuming 
a distance of 760 kpc. 

In most cases, we obtained acceptable fits using a power law spectral model 
(Table \ref{spec_par_GCS_xmm_powerlaw},\ref{spec_par_GCS_chandra_powerlaw}). However, for almost 
all brighter sources, a complex spectral models are required. For several sources with high 
luminosities (sources $\#\# 22, 26, 32, 42, 43$ in Table \ref{source_ID}), the models with 
quasi-exponential cut-off at $\sim 2 - 8$ keV or two-component models describe the energy spectra 
significantly better than a simple power law. To approximate the spectra of these sources, we used 
an absorbed power law model with exponential cut-off (XSPEC CUTOFFPL model), a Comptonization model 
and a two-component models.

For the Comptonization model approximation, we used the XSPEC model COMPTT (\cite{ST80,T94,TL95}). 
This model includes self-consistent calculation of the spectrum produced by the Comptonization 
of the soft photons in a hot plasma. It contains as free parameters the temperature of the 
Comptonizing electrons, $kT_{e}$, the plasma optical depth with respect to the electron scattering, 
$\tau$ and the temperature of the input Wien soft photon distribution, $kT_{0}$. A spherical 
geometry was assumed for the Comptonizing region.  

A cut-off power law and a Comptonization models gave reasonably good fit to the spectra of all sources 
considered, and the results are shown in Table \ref{spec_par_GCS_cutoffpl_comptt}. The model parameters 
derived for the M31 GC X-ray sources resemble those observed for bright Galactic low mass X-ray binaries 
(\cite{WSP88}; \cite{CS97}; \cite{CBC01}). 

The spectra of many luminous Galactic LMXB are well fit with a two-component model consisting of a 
soft black body-like component which might represent emission from an optically thick accretion disk 
or from the neutron star surface, together with hard component which may be interpreted as emission 
from a corona-like structure or a boundary layer between the disk and a neutron star (\cite{WSP88}; 
\cite{CS97}; \cite{Sidoli01}). We used such two-component models to approximate spectra of the three
brightest globular cluster sources showing spectral cut-off (sources $\#\# 22, 42, 43$). 
and $32$)   

For the soft component we used multicolor disk-blackbody (XSPEC DISKBB)(\cite{Mitsuda84}) 
and a black body (XSPEC BBODYRAD) models. DISKBB model has two parameters, the effective radius, 
$r_{in} \sqrt{cos i}$, where $r_{in}$ is the inner radius of the disk, $i$ is the inclination angle 
of the disk and $k T_{in}$ is the maximum color temperature of the disk. Although this model does 
not take into account the effects of electron scattering and uses simplified temperature profile in 
the accretion disk (\cite{SS73}), it has been found to give an adequate qualitative description to 
the spectra of accreting X-ray binaries (\cite{Mitsuda84}; \cite{WSP88}; \cite{CS97}). The BBODYRAD 
model has two parameters, the effective radius, $r_{BB}$ and color temperature, $k T_{BB}$. 

The hard spectral component can be adequately described by various phenomenological and physical 
models involving a break in the slope of the spectrum or quasi-exponential spectral cut-off 
at higher energies. For the sake of easier comparison with the results for the Galactic LMXB 
(\cite{WSP88}; \cite{CS97}; \cite{CBC01}), we use a simple black body BBODYRAD (in combination with 
DISKBB as a soft component) and a simple power law (in combination with BBODYRAD as a soft component) 
to approximate hard component in the spectra of brightest GC sources in our sample. The results of 
two-component approximation of the spectra of GC X-ray sources are shown in Table 
\ref{spec_par_GCS_two_comp}.

The two-component models give a good fit to the spectra of the high-luminosity GC X-ray sources in our 
sample. The characteristic temperature of the soft spectral component ranges from $0.5$ to $1.2$ 
keV, and the average contribution of the soft spectral component to the total source luminosity is 
$\sim 30\%$. Interestingly, even in the case of brightest M31 GC sources with total isotropic luminosities 
exceeding Eddington limit for the $1.4 M_{\odot}$ neutron star, the absorption-corrected luminosity of 
the soft spectral component never significantly exceeds that limit. In conclusion, the two-component 
approximation of the spectra of bright M31 GC sources gives spectral parameters consistent with high 
luminosity LMXBs hosting a neutron star (\cite{WSP88}; \cite{CS97}; \cite{CBC01}).

\subsection{Spectral distribution of GC X-ray sources}
\begin{figure}
\epsfxsize=18cm
\epsffile{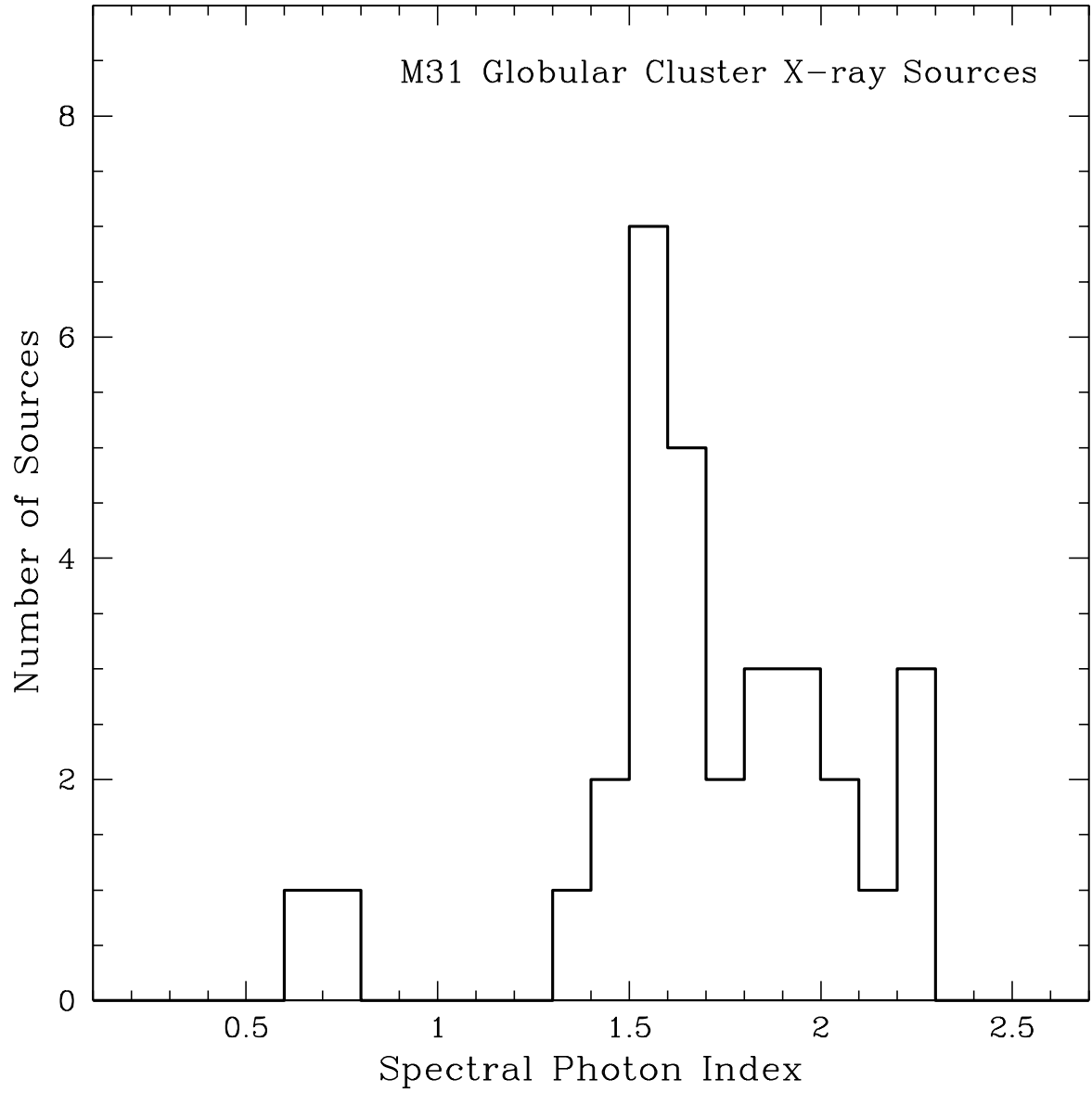}
\caption{The distribution of the spectral slopes derived from the analysis of 31 
bright GC X-ray sources. The histogram bins are $0.1$ wide. 
\label{hardness_distribution}}
\end{figure}

In order to characterize the overall spectral properties of the bright GC X-ray sources, we 
constructed a distribution of their spectral indices in the $0.3 - 7.0$ keV energy range using 
the model fits to both {\em XMM-Newton} and {\em Chandra} data with an absorbed simple power law 
(Fig. \ref{hardness_distribution}). For the sources with multiple spectral measurements, we used 
average values of the photon index. The main part of the distribution including 29 sources out of 
31 spans a range of photon indices between $\sim 1.4$ and $\sim 2.2$, and has a clearly defined 
maximum at $\alpha \sim 1.6$ (Fig. \ref{hardness_distribution}). Excluding two extremely hard 
X-ray sources, gives the corresponding weighted mean value of the photon index $1.65\pm0.01$.
This distribution is reminiscent of the spectral distribution of the X-ray sources in the central 
bulge of M31 (\cite{Shirey01}). The X-ray population of the central bulge of M31 is very likely to 
be dominated by bright LMXBs. Based on the similarity in the properties of M31 GC X-ray sources and 
their Galactic counterparts, we expect most of M31 GC X-ray sources to be LMXBs. Therefore the 
similarity between the spectral distributions of the central bulge of M31 and that of its GC X-ray 
sources is not surprising. Moreover, the peak of the M31 GC spectral index distribution coincides 
with the average value of the spectral photon index ($\alpha \sim 1.6$) derived for the LMXB 
populations in the nearby galaxies of different morphological type (\cite{Irwin03}). The spectral 
photon indices of the M31 GC X-ray sources in our sample were found to lie in the range, observed 
for the Galactic GC sources, in agreement with previous studies (\cite{DiStefano01}).

\subsection{Hardness-luminosity diagram of GC X-ray sources}
\begin{figure}
\epsfxsize=18cm
\epsffile{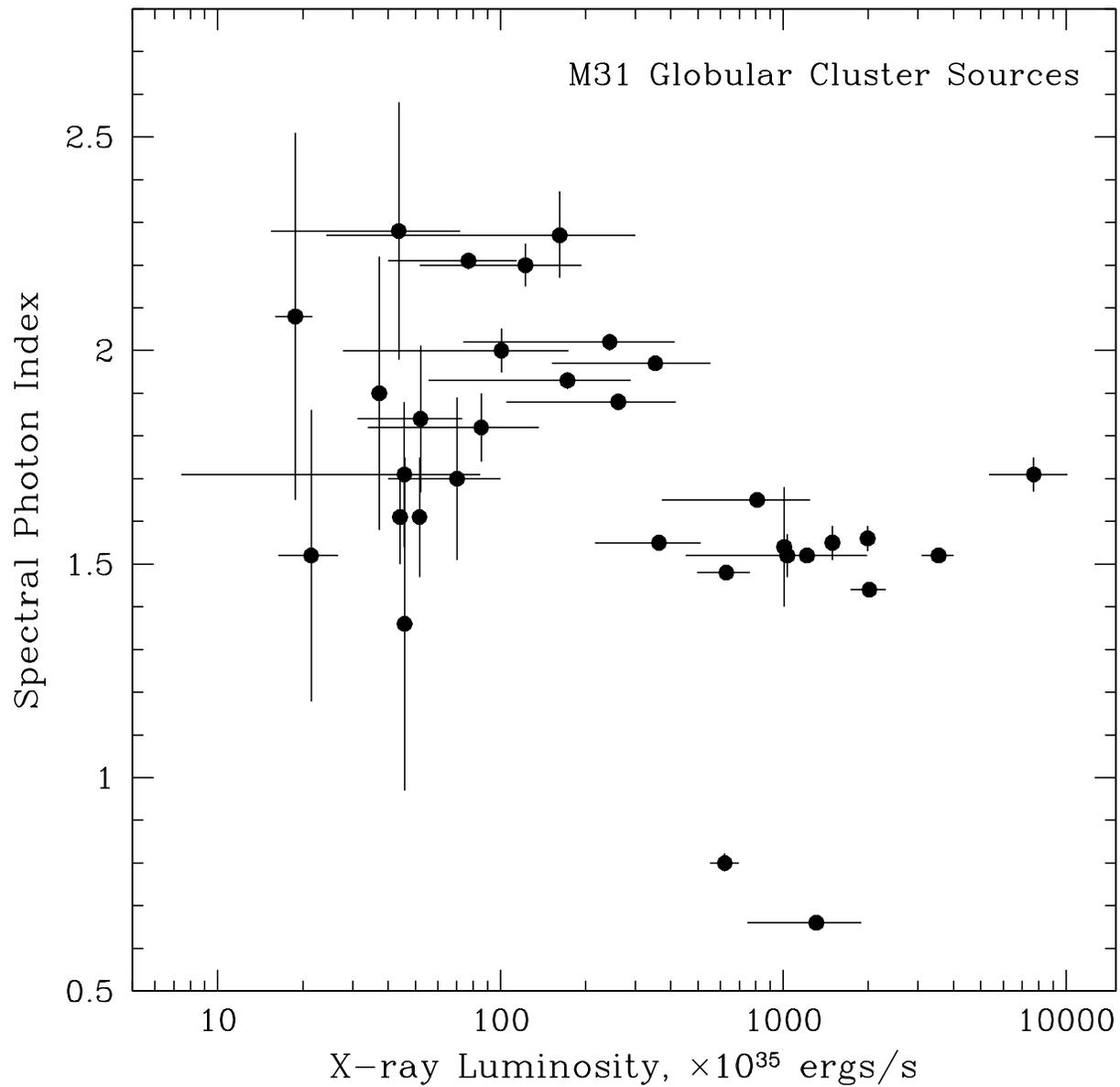}
\caption{Spectral photon index of the GC X-ray sources in M31 vs. their X-ray 
luminosity in the $0.3 - 10$ keV energy band. The error bars in X-axis reflect 
statistical uncertainty of the source flux determination and in some cases the 
range of source X-ray luminosities observed with {\em XMM} and {\em Chandra}. 
\label{hardness_luminosity}}
\end{figure}

We studied the relation between the hardness of the spectrum and luminosity of GC X-ray sources in 
our sample. In Fig. \ref{hardness_luminosity} the hardness of the spectrum of GC X-ray sources 
expressed in terms of spectral photon index is shown as a function of their X-ray luminosity in the 
$0.3 - 10$ keV energy band. As it is clearly seen from Fig. \ref{hardness_luminosity}, there are 
three groups of sources, occupying distinct regions in the hardness-luminosity diagram. The first 
group represents majority of the brightest X-ray sources with luminosities above 
$\sim 4 \times 10^{37}$ ergs s$^{-1}$. The objects in this group have relatively narrow range of 
photon indices with average value of $1.55$, suggesting common spectral formation mechanism at high 
luminosities. The second group comprised of fainter sources, shows greater variety of photon indices 
ranging from $\sim 1.4$ to $\sim 2.3$. Based on the wide range of X-ray luminosities, this group could 
represent a mix of LMXBs in both low and high luminosity states (\cite{WSP88}). The third group consists 
of two objects with extremely hard spectra ($\alpha \sim 0.6 - 0.8$). Both objects also have a more 
pronounced spectral cut-off than the rest of the GC sources (Table \ref{spec_par_GCS_cutoffpl_comptt}). 
In addition, one of the sources, the X-ray source $\# 32$ in the globular cluster Bo 158, shows regular 
X-ray dips, implying a system observed at high inclination angle (\cite{dipper02}). The 
hardness-luminosity diagram of GC X-ray sources in our sample appears to be consistent with that of the 
central bulge (\cite{Shirey01}).      

\subsection{Low energy absorption}
The unprecedented sensitivity of {\em XMM} and {\em Chandra} observations allows a reliable 
measurement of low energy absorption toward brighter sources in M31. Using the results of 
spectral fitting, we have been able to estimate the value of equivalent hydrogen absorbing 
column, $N_{\rm H}$ for 31 source in our sample. To make a homogeneous sample, we used the 
same simple spectral model to estimate $N_{\rm H}$ for all sources (an absorbed power law model).
It should be noted, that the amount of low-energy absorption derived from spectral fitting 
sometimes depends strongly on the type of continuum model used to approximate X-ray spectrum 
(Table \ref{spec_par_GCS_xmm_powerlaw},\ref{spec_par_GCS_cutoffpl_comptt}). The simple power law 
approximation does not take into account possible presence of an additional soft X-ray component 
leading to the underestimation of the absorbing column. The choice of a particular model for the 
soft X-ray component (i.e. black body vs. multicolor disk black body) or a hard spectral continuum 
(i.e. a Comptonization model vs. simple power law) introduces additional uncertainty in the value 
of $N_{\rm H}$.  
  
Our analysis shows that for the large fraction of GC sources the derived value of $N_{\rm H}$ 
is either in excess or consistent with Galactic hydrogen column 
$N_{\rm H}^{Gal} \sim 7 \times 10^{20}$ cm$^{-2}$ in the direction of M31 (\cite{DL90}). In 
the first case, it is natural to assume that additional absorption could be due to interstellar 
matter outside our Galaxy (i.e. in M31 and intergalactic medium) and/or within the X-ray binary 
itself. We compared the values of $N_{\rm H}$, derived from our spectral fits with values 
predicted using the optical observations. If there is no extra absorption within the X-ray 
binary itself, a close agreement between the absorbing column values derived from analysis of 
X-ray and optical data is expected (\cite{Sidoli01}). 

\begin{figure}
\hbox{
\begin{minipage}{9.0cm}
\epsfxsize=9.0cm
\epsffile{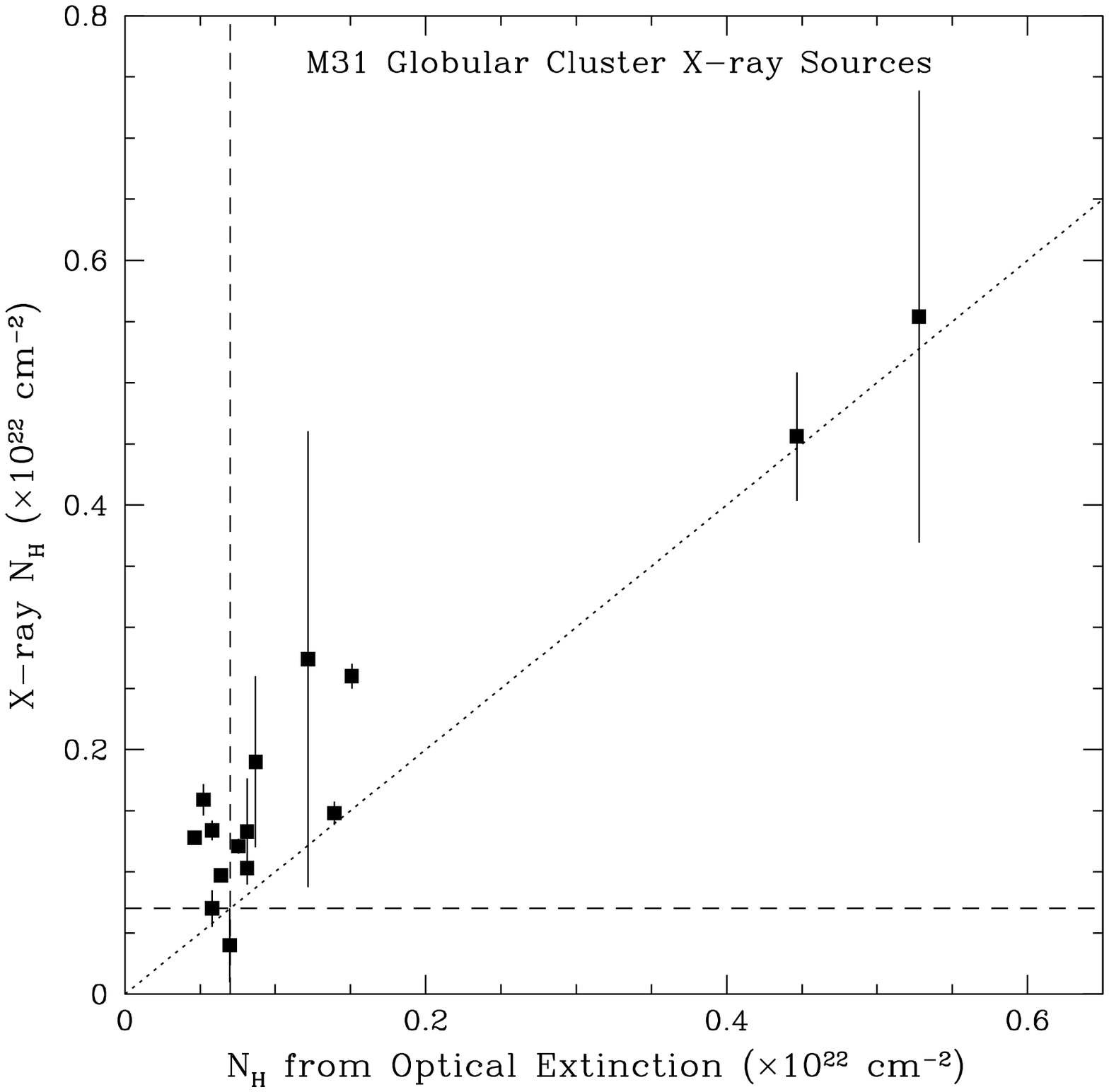}
\end{minipage}
\begin{minipage}{9.0cm}
\epsfxsize=9.0cm
\epsffile{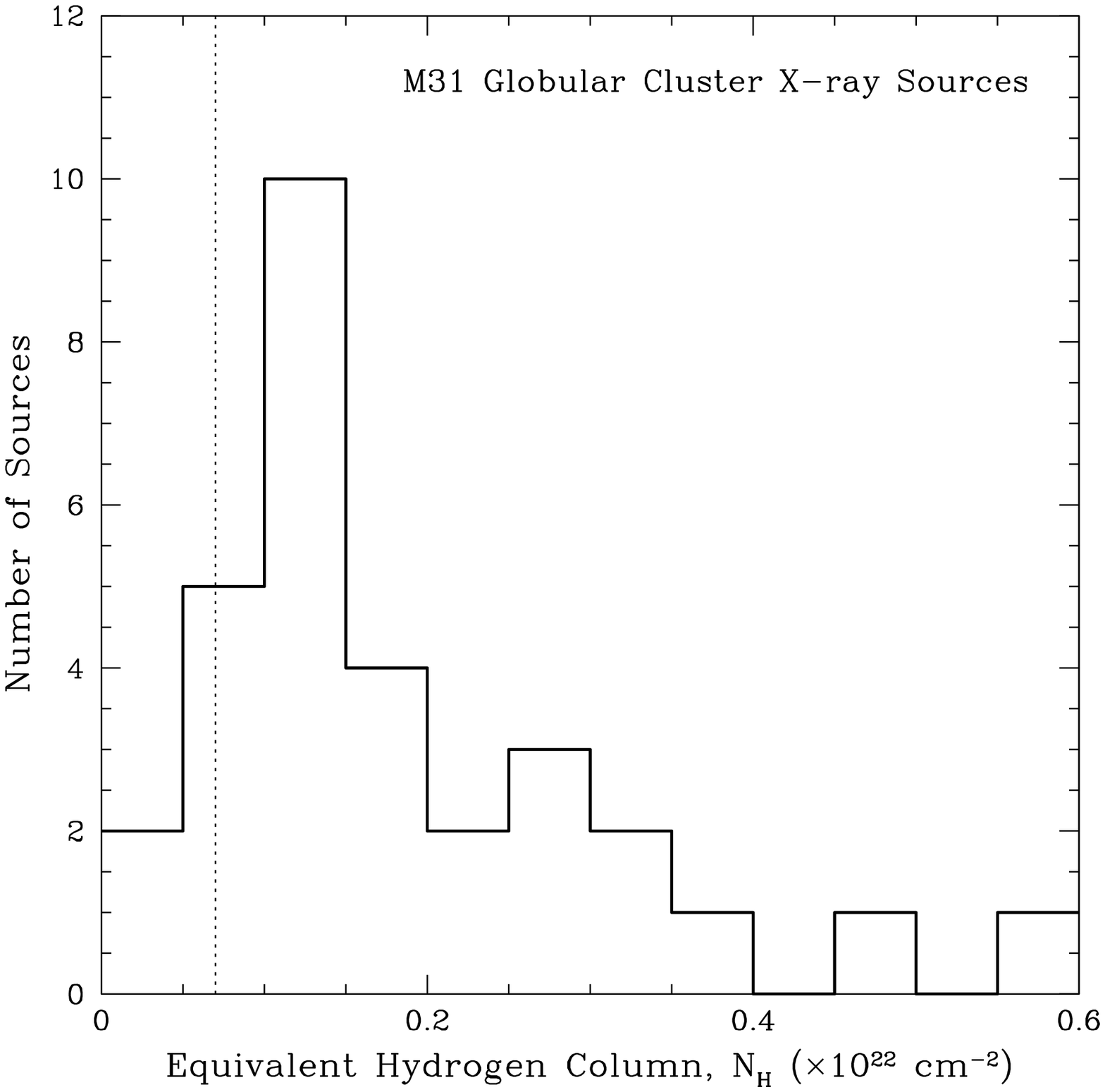}
\end{minipage}
}
\caption{{\em Left panel:} The values of $N{\rm H}$ derived from X-ray spectral 
fitting plotted against the values of $N_{\rm H}$ derived from optical extinction 
data. The dotted line shows points with equal X-ray and optical $N_{\rm H}$. The 
dashed lines mark expected Galactic foreground absorbing column in the direction 
of M31. {\em Right panel:} The distribution of the absorbing columns derived from 
the spectral analysis of 31 bright GC X-ray sources. Each bin along X-axis has a 
width of $5\times 10^{20}$ cm$^{-2}$. The Galactic foreground absorbing column in 
the direction of M31 ($7 \times 10^{20}$ cm$^{-2}$) is marked with dotted line. 
\label{N_H_opt_x_ray}}
\end{figure}

Using the results of optical observations of M31 globular clusters 
(\cite{Barmby00}; \cite{Huchra91}; \cite{Perrett02}), we were able to estimate the values of optical 
extinction for 15 GC sources with measured X-ray low-energy absorption. The values of optical color 
excess, $E_{B-V}$ were converted to the equivalent hydrogen column densities, $N_{\rm H}^{\rm opt}$ 
using the relation: $N_{\rm H}/E_{\rm B-V} = 5.9 \times 10^{21}$ cm$^{-2}$ mag$^{-1}$ (\cite{cox00}). 
In Fig. \ref{N_H_opt_x_ray} ({\em left panel}) the values of the X-ray low-energy absorption, 
$N_{\rm H}^{\rm X}$ are plotted against optically determined values, $N_{\rm H}^{\rm opt}$. The dotted 
line corresponds to equal X-ray and optically determined absorption. Taking the aforementioned systematic 
uncertainties in the derivation of X-ray absorption into account there is a general agreement between 
two absorption measurements (Fig. \ref{N_H_opt_x_ray}, {\em left panel}). Based on the current 
observations it is impossible to confirm or rule out a possibility of additional absorption in some 
sources. More sensitive observations with wider bandpass (i.e. future Constellation-X) and physically 
justified detailed spectral modeling are needed to resolve this interesting presently open issue. 

The distribution of the X-ray absorption could provide additional information on the geometry 
of M31 globular cluster system and help to map the structure of ISM inside M31. For example, the 
sources located behind or inside the disk of M31 should be in general more absorbed/reddened than 
sources seen in front of the disk. We constructed the distribution of absorption columns for 31 
M31 GC sources (Fig. \ref{N_H_opt_x_ray}, {\em right panel}). The distribution has a prominent 
peak centered at $\sim 1.2 \times 10^{21}$ cm$^{-2}$ and a tail structure extending up to 
$\sim 6 \times 10^{21}$ cm$^{-2}$. In general, the observed $N_{\rm H}$ distribution could be 
consistent with homogeneous quasi-spherical spatial distribution of GC sources with an additional 
absorption from the disk.

\section{VARIABILITY}

\subsection{Long-term variability}
\begin{figure}
\hbox{
\begin{minipage}{9.0cm}
\epsfxsize=9.0cm
\epsffile{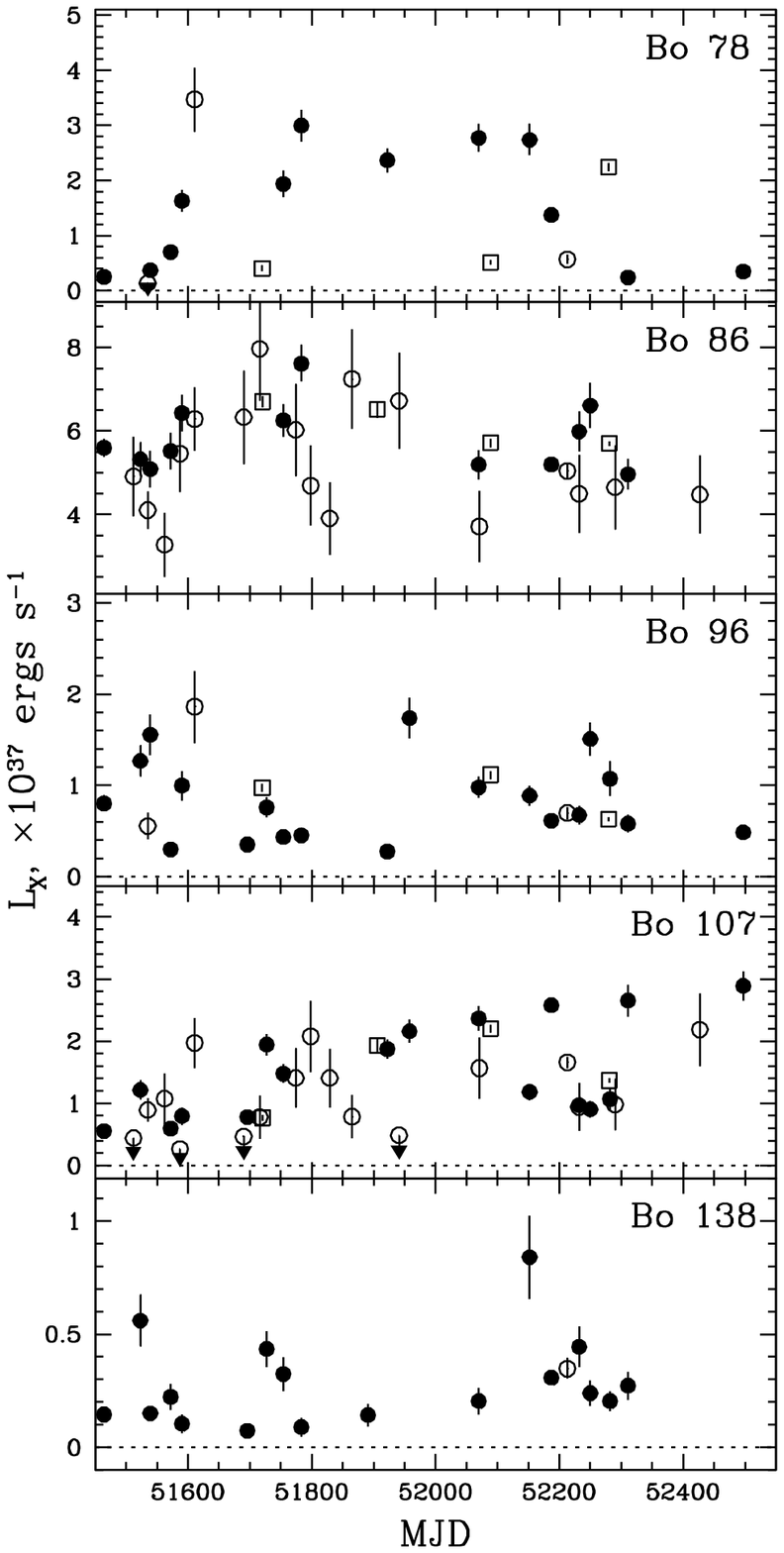}
\end{minipage}
\begin{minipage}{9.0cm}
\epsfxsize=9.0cm
\epsffile{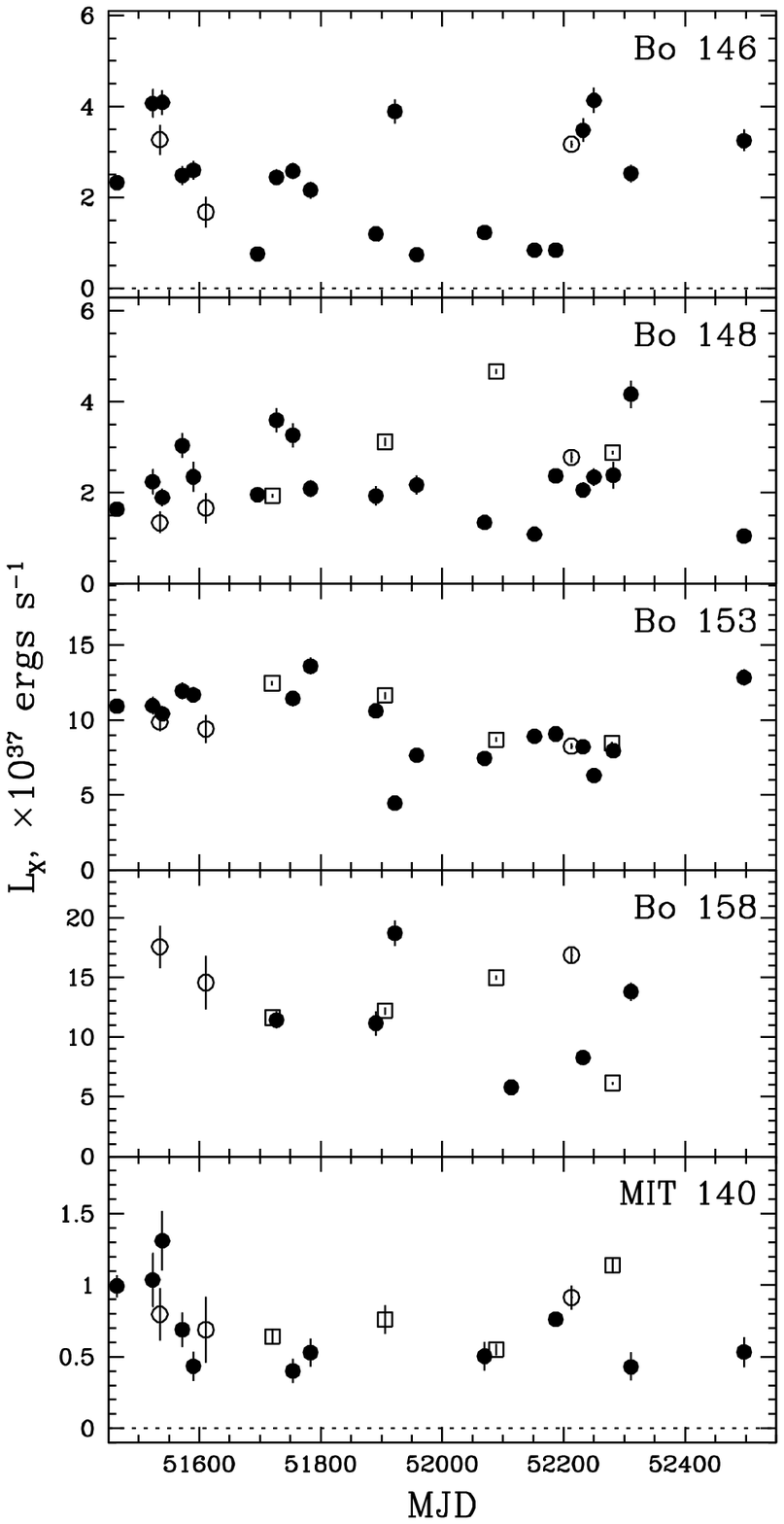}
\end{minipage}
}
\caption{X-ray flux histories of 10 GC sources in our sample obtained combining 
the data of {\em Chandra}/ACIS ({\em filled circles}), HRC ({\em open circles}) 
and {\em XMM}/EPIC ({\em open rectangles}) observations, $0.3 - 10$ keV energy 
range. The upper limits correspond to a $2 \sigma$ level. 
\label{long_term_var}}
\end{figure}

We searched for the long-term flux variability of M31 GC sources in our sample combining the 
data of multiple {\em Chandra} and {\em XMM-Newton} observations. We found that more than 
$80\%$ of sources with multiple flux measurements available show significant variability 
(Table \ref{source_ID}). In Fig. \ref{long_term_var} the long-term flux histories of ten M31 GC 
sources demonstrating the highest levels of variability are shown. These sources show both 
irregular and possible quasi-periodic patterns of X-ray variability with corresponding flux changes 
by more than a factor of $2$ over a time scales of several days to years.

The persistent low luminosity GC X-ray sources with average luminosities below $few \times 10^{37}$ 
ergs s$^{-1}$ tend to be generally more variable than brighter ones. The corresponding ratio between 
the highest and the lowest flux levels observed with {\em XMM} and {\em Chandra} is between $\sim 3$ 
and $> 15$ for low luminosity sources, while it is usually lower than $\sim 3$ for the high luminosity 
sources.   

The X-ray source associated with globular cluster Bo 107 shows a possible pattern of recurrent 
outbursts lasting for $\sim 50 - 100$ days spaced by $\sim 100 - 200$ day intervals. The source 
X-ray flux during the peak of the outburst can be at least $\sim 10$ times higher than during the 
low luminosity periods. This type of variability is reminiscent of the Galactic LMXB like Aql X-1 
(\cite{PT84}).

Di Stefano et al. (2002) proposed that the X-ray source in Bo 86 may have long-term periodicity on 
a time scale of $\sim 200$ days (see also Kong et al. 2002). Our data, though covering a longer 
period of time, does not allow us to confirm or completely rule out that kind of variability of Bo 
86. More frequent observations would be needed to establish variability of the source.

\begin{figure}
\hbox{
\begin{minipage}{9.0cm}
\epsfxsize=9.0cm
\epsffile{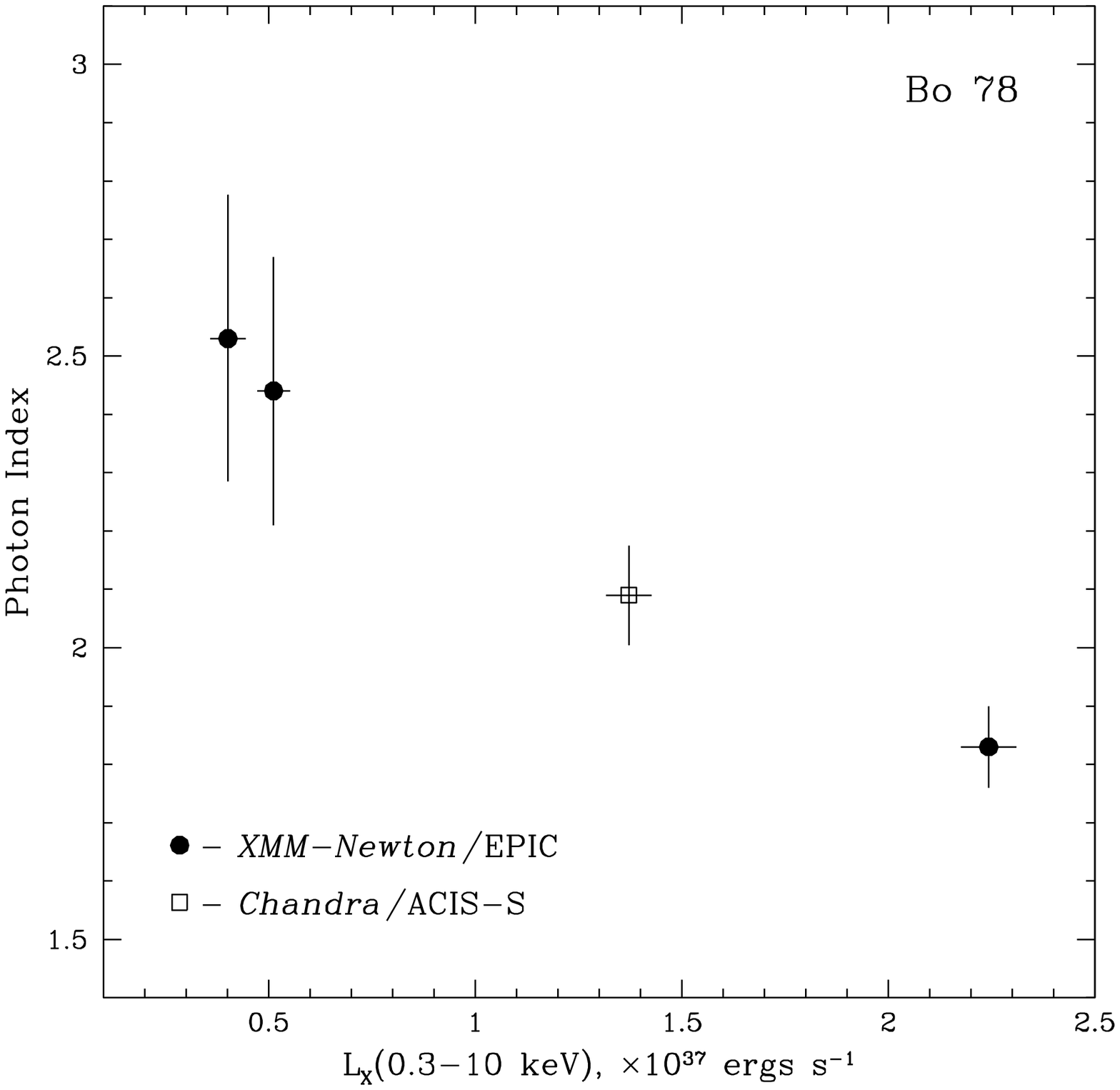}
\end{minipage}
\begin{minipage}{9.0cm}
\epsfxsize=9.0cm
\epsffile{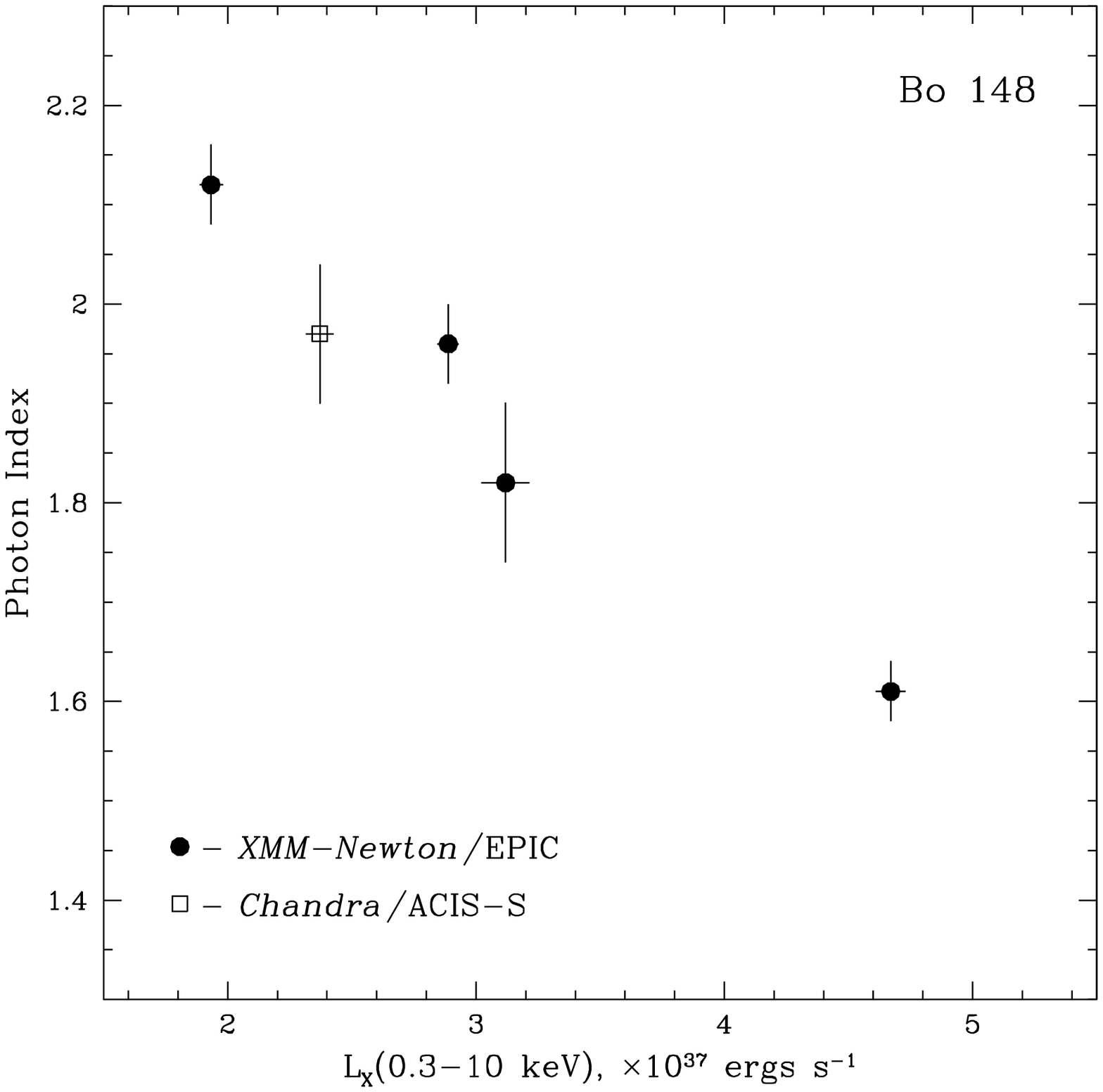}
\end{minipage}
}
\caption{Spectral variability of M31 GC sources Bo 78 ($\# 9$)({\em left panel}) and 
Bo 148 ($\# 29$) ({\em right panel}). The hardness of the spectrum expressed in terms 
of the spectral photon index (Y-axis) is plotted against source X-ray luminosity in 
the $0.3 - 10$ keV energy band (X-axis). \label{spec_var}}
\end{figure}

We also searched for the long-term spectral variability of all bright M31 GC sources using the 
results of spectral analysis of {\em XMM} and {\em Chandra} data. Several sources demonstrate 
a correlation between the level of X-ray flux and hardness of their energy spectrum. The 
spectral fits for most of variable sources indicate that as the source flux increases, the 
spectrum becomes harder (Table \ref{spec_par_GCS_xmm_powerlaw}). Two characteristic examples 
of such behavior are shown in Fig. \ref{spec_var} (sources Bo 78 and Bo 148). In this figure 
the hardness of X-ray spectrum expressed in terms of best-fit photon index is shown as a function 
of the source luminosity. The luminosity of the spectrally variable sources ranges from 
$\sim 10^{36}$ to $\sim few \times 10^{38}$ ergs s$^{-1}$, falling into typical range of 
luminosities for both atoll and Z-source classes of Galactic LMXB (\cite{HK89}). For brighter 
sources with luminosities around $10^{38}$ ergs s$^{-1}$ the observed spectral changes could be 
consistent with that of the Z-sources on the normal branch (\cite{HK89}).

\subsection{Recurrent transient activity of Bo 163.}
\begin{figure}
\epsfxsize=18cm
\epsffile{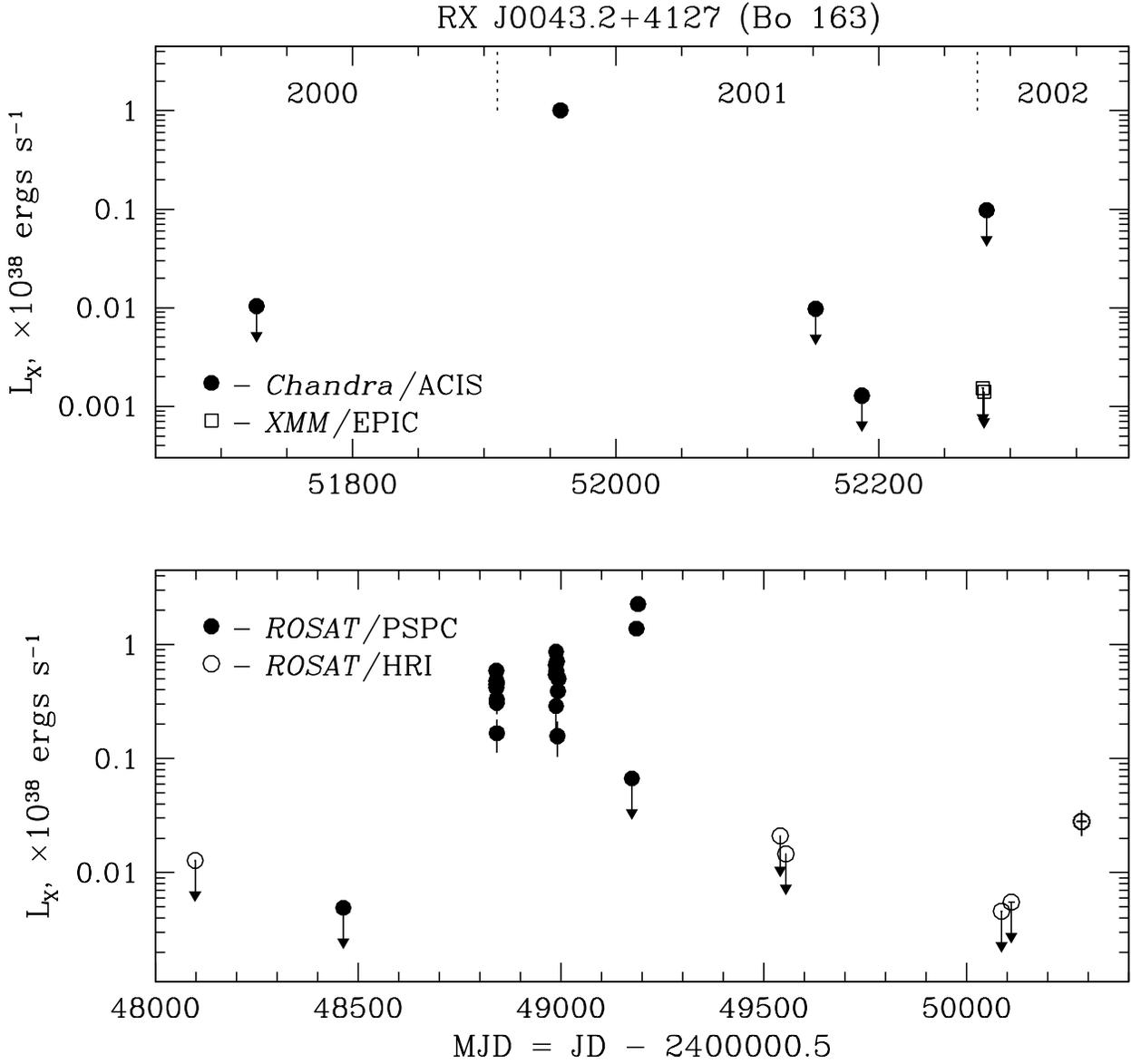}
\caption{X-ray flux histories of the recurrent M31 GC source Bo 163 ($\# 35$) 
based on {\em XMM-Newton} and {\em Chandra} ({\em upper panel}) and {\em ROSAT} 
({\em lower panel}) archival data. The upper limits correspond to a $2 \sigma$ 
level. 
\label{GCS_N_CH_003_lc}}
\end{figure}

One of the sources, coincident with a globular cluster Bo 163 shows a recurrent 
transient-like variability. Our analysis of the archival {\em Chandra} data revealed 
the presence of a bright X-ray source CXO J004317.8+412745 with a position consistent 
with Bo 163 and previously detected {\em ROSAT} source RX J0043.2+4127 (\cite{Supper01}) 
during 2001 February 18 ACIS-S observation (Obs. ID $\#1582$) (Fig. \ref{GCS_N_CH_003_lc}, 
{\em upper panel}). The estimated X-ray luminosity of the source was $\sim 10^{38}$ 
ergs s$^{-1}$ in the $0.3 - 10$ keV energy band (see source spectral parameters in 
Table \ref{spec_par_GCS_chandra_powerlaw}). Other {\em XMM-Newton} and {\em Chandra} 
observations covering the position of Bo 163 did not yield positive source detection with 
$2 \sigma$ upper limits on its luminosity ranging from $\sim 10^{35}$ to $\sim 10^{37}$ 
ergs s$^{-1}$ (depending on the type of instrument and exposure time). 

We analyzed the data of the earlier observations of Bo 163/RX J0043.2+4127 with {\em ROSAT}/PSPC 
and HRI detectors. The observed {\em ROSAT} source count rates were converted into the X-ray 
fluxes in the $0.3 - 10$ keV energy band using WebPIMMS assuming an absorbed power law 
spectrum with $N_{\rm H} = 7 \times 10^{20}$ cm$^{-2}$ and $\alpha = 1.7$. The resulting 
{\em ROSAT} light curve of RX J0043.2+4127 is shown in {\em lower panel} of 
Fig. \ref{GCS_N_CH_003_lc}. The source clearly demonstrates a pattern of recurrent 
transient-like behavior with bright outbursts with $L_{\rm X} > 10^{38}$ ergs s$^{-1}$ 
and an extended quiescent periods with $L_{\rm X} < 10^{36}$ ergs s$^{-1}$. The corresponding 
maximum ratio of the the highest outburst luminosity ($L_{\rm X} \sim 2 \times 10^{38}$ ergs 
s$^{-1}$) to the lowest measured quiescent luminosity ($L_{\rm X} < 10^{35}$ ergs s$^{-1}$) 
is $> 2000$. 

The overall X-ray properties of RX J0043.2+4127 are reminiscent of both Galactic neutron star 
and black hole recurrent transients like Cen X-4 (\cite{KHS80,Paradijs87}) and 4U1630-40 
(\cite{Priedhorsky86}). Unfortunately, current spectral and timing data does not allow us to 
determine the nature of the compact object in RX J0043.2+4127. Observations with higher 
sensitivity and timing resolution are needed to choose between neutron star and black hole 
interpretation.  

\subsection{Short-term variability.}
The comparisons based on the broad-band spectral properties are not sufficient to establish 
a neutron star nature or rule out a black hole nature for the M31 globular cluster sources. 
On the other hand, the study of their short-term variability can provide a definitive answer, 
if Type I X-ray bursts or X-ray pulsations are observed. We searched for both types of 
variability in the M31 GC data. Unfortunately, the observed source count rates for most sources 
are too low to allow us to probe timescales shorter than $\sim 20 - 30$ seconds. We did not find 
evidence of X-ray pulsations and short X-ray bursts in the {\em XMM-Newton} and {\em Chandra} 
data. 

\begin{figure}
\epsfxsize=18cm
\epsffile{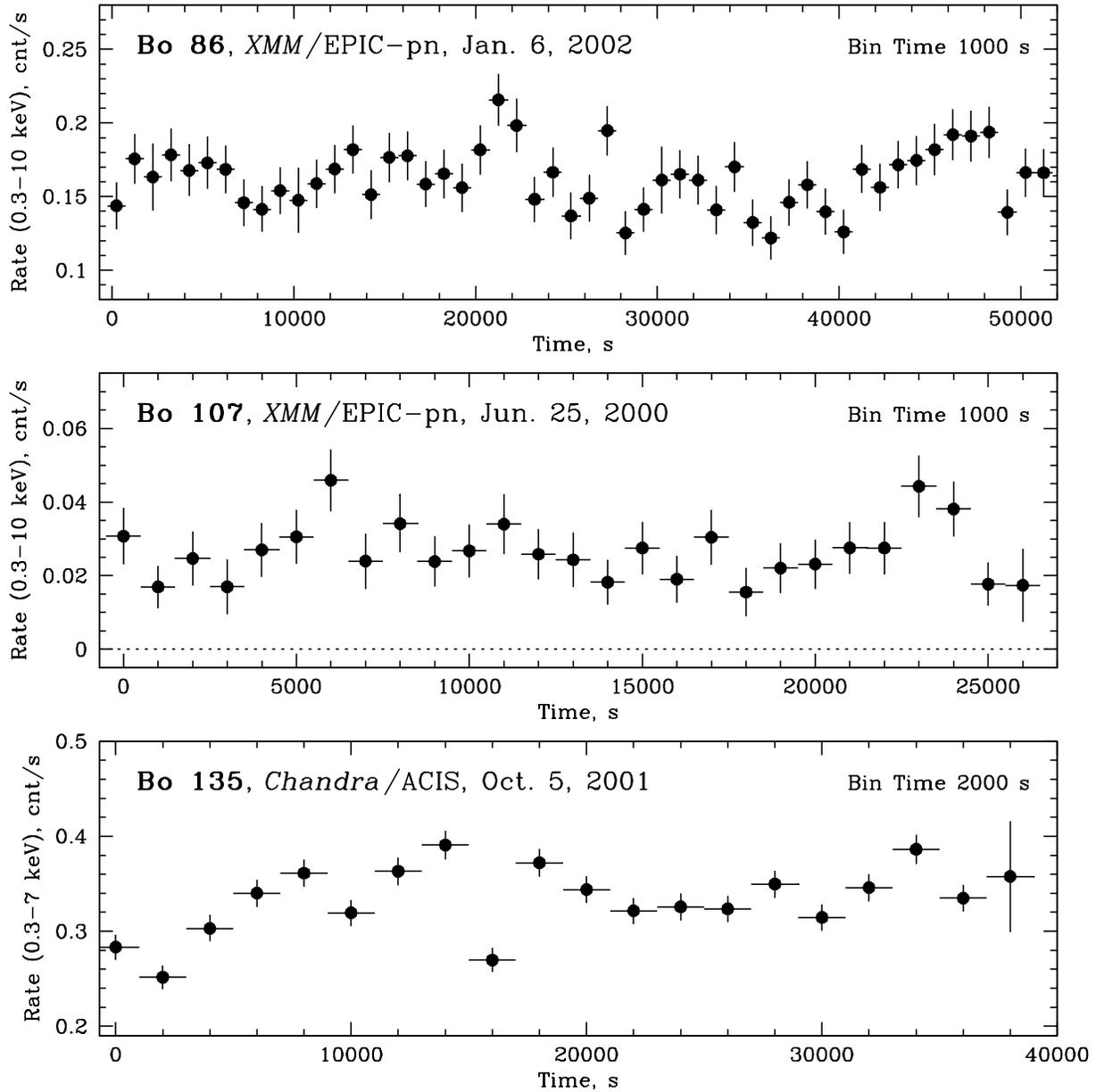}
\caption{{\em Upper panel:} X-ray light curve of Bo 86 ($\# 11$) during the 2002 
January 6 {\em XMM-Newton}/EPIC-pn observation ($0.3 - 10$ keV energy range and a 
$1000$ s time resolution). {\em Middle panel:} X-ray light curve of Bo 107 ($\# 16$) 
during the 2000 June 25 {\em XMM-Newton}/EPIC-pn ($0.3 - 10$ keV energy range and a 
$1000$ s time resolution). {\em Lower panel:} X-ray light curve of Bo 135 ($\# 22$) 
during the 2001 October 5 {\em Chandra}/ACIS observation ($0.3 - 7$ keV energy range 
and a $2000$ s time resolution). \label{short_term_lc}}
\end{figure}

Several GC sources demonstrate significant aperiodic variability on a time scale of individual 
observation. As an example of such behavior, the lightcurves of three GC sources are shown 
in Fig. \ref{short_term_lc}. All three sources (Bo 86, Bo 107 and Bo 135) show irregular changes 
of X-ray flux of up to $\sim 100 \%$ over a time scales of $1000 - 10000$ s. These sources 
were also found to be variable on a time scales of months to years (Fig. \ref{long_term_var}). 
The combination of such short and long term variabilities was long recognized as one of the 
characteristic features of X-ray binaries (\cite{HK89}). In addition, the X-ray source coincident 
with M31 GC candidate Bo 158 shows a remarkable dip-like modulation with a a period of $10017$ s 
(\cite{dipper02}).

\begin{figure}
\epsfxsize=18cm
\epsffile{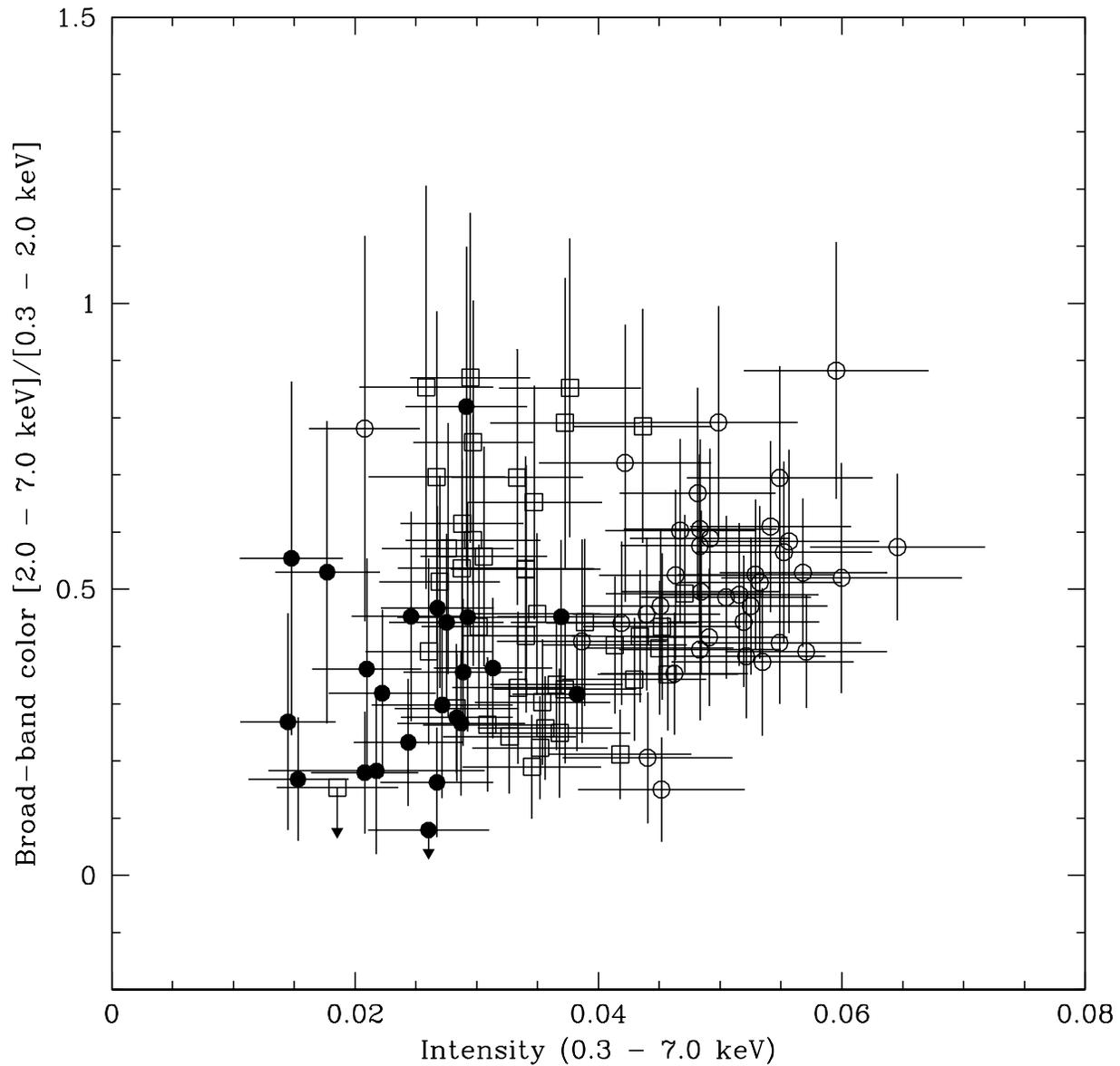}
\caption{Broad-band color (hardness) vs. intensity for three {\em XMM-Newton} 
observations of X-ray source in the globular cluster Bo 148 ($\# 29$). The broad-band 
color is defined as ratio of source intensities in the $2.0 - 7.0$ and $2 - 7$ keV 
energy bands. The EPIC-MOS data is binned to 1500 s. The corresponding source luminosity 
changes between $\sim 10^{37}$ and $\sim 4 \times 10^{37}$ ergs s$^{-1}$ in the $0.3 - 10$ 
keV energy band. 
\label{hardness_int_diagram}}
\end{figure}

The unprecedented sensitivity of {\em XMM-Newton} provides a unique opportunity to study spectral 
evolution of the brightest LMXB sources in M31 on a time scale of hundreds and thousands of seconds. 
To study short-term spectral variability of the bright GC sources, we constructed their X-ray 
hardness-intensity diagrams using the data of {\em XMM-Newton} observations. The X-ray hardness was 
defined as the ratio of the source intensities in the $2.0 - 7.0$ keV and $0.3 - 2$ keV energy bands 
with data integration times of $1000 - 3000$ s depending on the source intensity. Several sources 
show a complex patterns of evolution on the hardness-intensity diagram somewhat reminiscent of the 
Galactic Z and atoll sources (\cite{HK89}). In Fig. \ref{hardness_int_diagram}, we show the 
hardness-intensity diagram of one of these sources, the X-ray source in globular cluster Bo 148.
Although the data strongly suggest some pattern of spectral evolution, it is still insufficient to 
classify source behavior in conventional framework of Galactic LMXB. More high-sensitivity 
observations covering wider range of source luminosities are needed to address this issue. 

\subsection{``Dipping'' X-ray source in Bo 158}
\begin{figure}
\epsfxsize=18cm
\epsffile{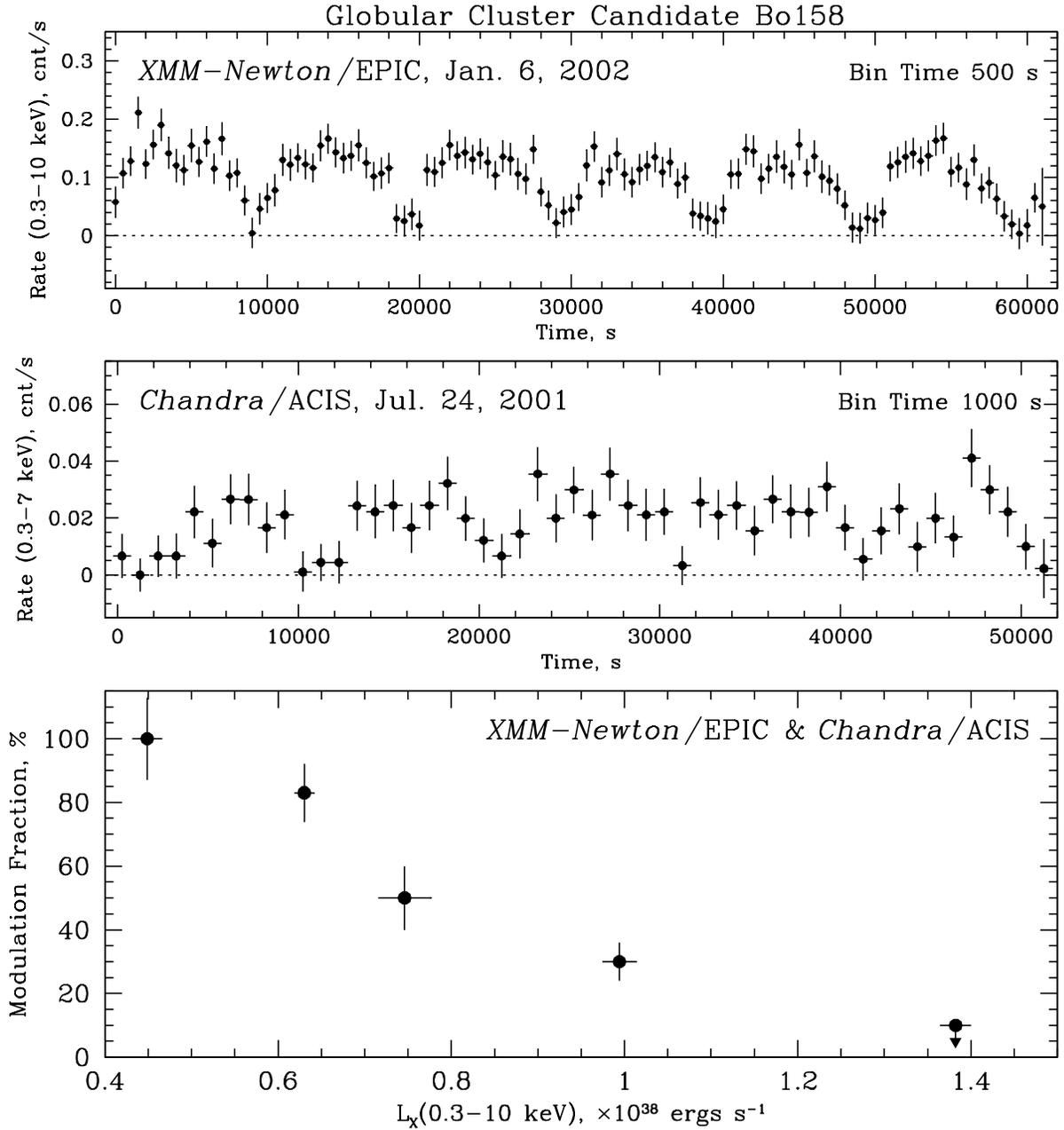}
\caption{{\em Upper panel:} X-ray light curve of Bo 158 ($\# 32$) during the 2002 
January 6 {\em XMM-Newton} observation, obtained from combined EPIC-pn, MOS1, and MOS2 
cameras ($0.3 - 10$ keV energy range and a $500$ s time resolution). {\em Middle panel:} 
X-ray light curve of Bo 158 during the 2001 July 24 {\em Chandra}/ACIS observation 
($0.3 - 7$ keV energy range and a $1000$ s time resolution). {\em Lower panel:} 
The X-ray modulation fraction ($\%$) plotted as a function of X-ray luminosity of 
Bo 158 (in units of $10^{38}$ ergs s$^{-1}$) \label{dipper_lc}}
\end{figure}

The X-ray source J004314.1+410724 was discovered in M31 by the {\em Einstein} observatory 
(source $\#85$ in \cite{TF91}) and was detected in subsequent observations with {\em ROSAT} 
(\cite{Primini93,Supper01}), {\em XMM-Newton} (\cite{dipper02}) and {\em Chandra} 
(\cite{DiStefano01}). The position of the source is consistent with the optically identified 
globular cluster candidate Bo 158 (source $\#158$ in Table IV of \cite{Bo87}). The observation 
of the central bulge of M31 with {\em XMM-Newton} taken 2002 Jan 6 revealed strong periodic 
dip-like modulation of the source flux (\cite{dipper02}; Fig. \ref{dipper_lc}, {\em upper panel}). 
The X-ray flux was modulated by $\sim 83\%$ at a period of 2.78 hr (10017 s). The X-ray intensity 
dips show no energy dependence. The analysis of earlier archival {\em XMM} and {\em ROSAT} 
observations taken in the year 2000 and 1991 revealed weaker dips with amplitudes of $\sim 30 \%$ 
and $\sim 50 \%$. The amplitude of the modulation has been found to be anticorrelated with 
source X-ray flux. A detailed description of the timing and spectral properties 
of Bo 158 are presented in Trudolyubov et al. 2002{\em b}.

We extended our original study of the X-ray modulation in Bo 158 combining the data of 
archival {\em XMM} and {\em Chandra} observations. Because of a relatively long period of 
modulation ($\sim 10000$ s), only one {\em Chandra} observation was selected for our analysis 
(Obs. $\# 2017$ with ACIS). For one {\em XMM} observation (Obs. $\# 1$), showing significant 
change of the source flux on a time scale of $\sim 5$ hours, we divided the data into two 
segments corresponding to different luminosity levels. A sample light curves of Bo 158 obtained 
with {\em XMM} and {\em Chandra} are shown in the upper and middle panels of Fig. \ref{dipper_lc}. 
The modulation fraction (i.e. the ratio of the average modulation amplitude to the average source 
flux) was determined for each observation. The resulting dependence of the modulation fraction on 
the source flux is shown in the bottom panel of Fig. \ref{dipper_lc}. The relative amplitude of 
X-ray modulation declines dramatically (at least by factor of $\sim 10$) as the source 
luminosity rises from $\sim 4.5 \times 10^{37}$ to $\sim 1.4 \times 10^{38}$ ergs s$^{-1}$. The 
observed correlation between the modulation fraction and X-ray flux of Bo 158 allows us to put 
constraint on possible number of bright X-ray sources presently active in this cluster. We 
conclude that one X-ray source must contribute up to $99\%$ of the detected flux.  

In addition to the short-term periodic variability, there is clear evidence for significant 
long-term variability of Bo 158. The long-term monitoring observations with {\em Chandra} and 
{\em XMM-Newton} (Fig. \ref{long_term_var}) show that the source is highly variable on a time 
scales of months to years. The X-ray luminosity Bo 158 changes at least by factor of $\sim 4$, 
e.g. between $\sim 5 \times 10^{37}$ and  $\sim 2 \times 10^{38}$ ergs s$^{-1}$.

\section{Luminosity distribution of M31 GC X-ray sources.}
\begin{figure}
\epsfxsize=18cm
\epsffile{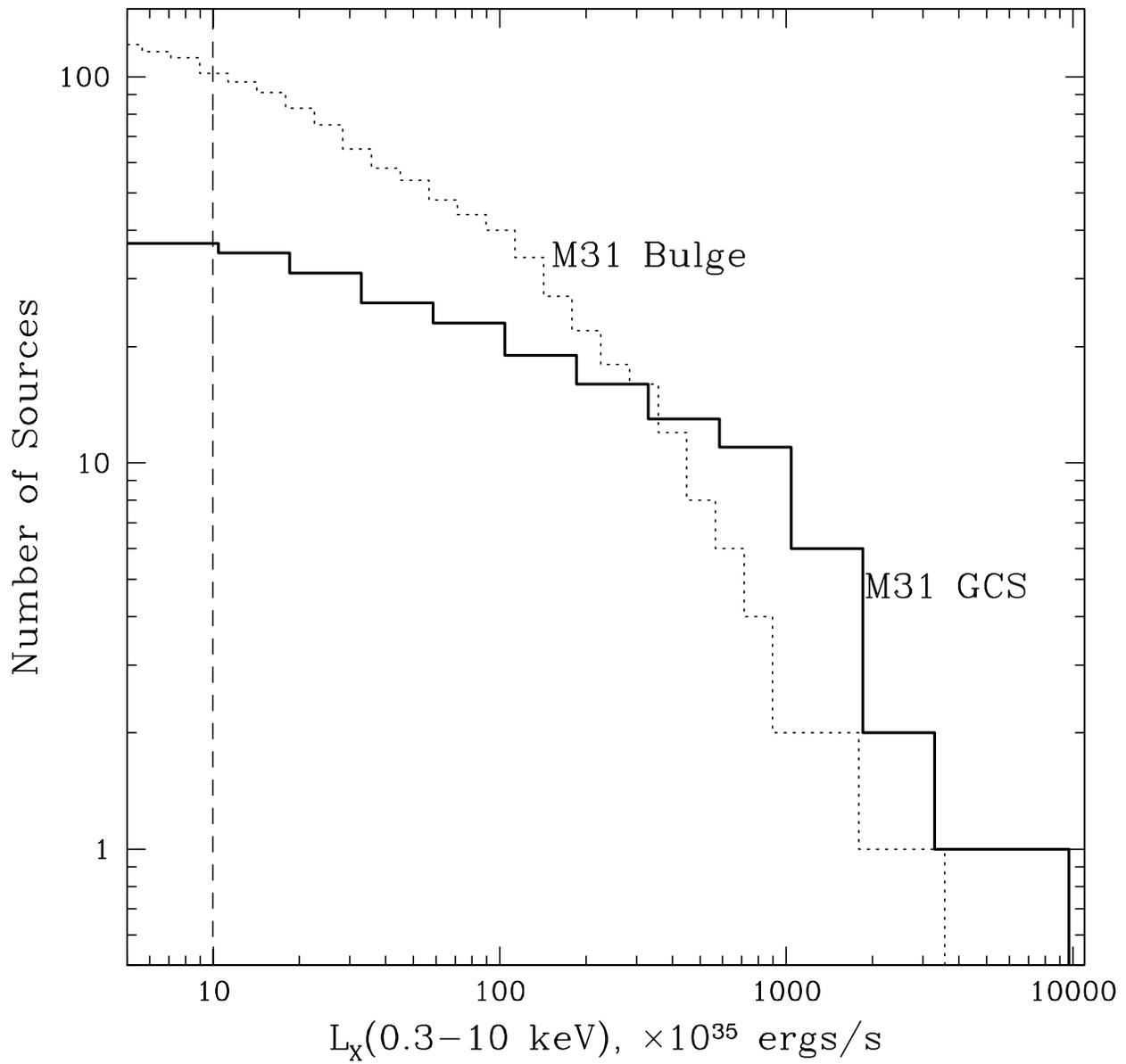}
\caption{Cumulative X-ray luminosity distributions of M31 GC X-ray sources and central 
bulge of M31.\label{GCS_XLF}}
\end{figure}

We have built differential and cumulative luminosity distributions of M31 GC X-ray 
sources assuming a distance of 760 kpc (Fig. \ref{GCS_XLF}). The luminosity 
distribution for the central bulge based on the results of {\em XMM-Newton} observations 
(Trudolyubov et al. 2002{\em a}) is also shown in Fig. \ref{GCS_XLF}. Luminosities of the 
brighter sources ($L_{\rm X} > 5 \times 10^{36}$ ergs s$^{-1}$) were derived using analytical 
fits to their spectra in the $0.3 - 10$ keV energy band. For faint sources, luminosities 
were derived using WebPIMMS, assuming an absorbed power law model with 
$N_{\rm H} = 7 \times 10^{20}$ cm$^{-2}$ (\cite{DL90}) and $\alpha = 1.7$ (\cite{Shirey01}) 
in the same energy band. For sources with multiple flux measurements available, we used 
average flux to calculate the luminosity. Taking into account the sensitivity as a function of 
off-axis distance, different exposure times and the effect of diffuse X-ray emission near the 
center of M31, we estimate flux completeness limit of our samples as $\sim 10^{36}$ ergs 
s$^{-1}$ (indicated by dotted line in Fig. \ref{GCS_XLF}).

The shapes of both cumulative and differential luminosity distributions for M31 GC 
sources indicate the presence of a cut-off and can be described by a broken power law 
or a cut-off power law models. We fitted the cumulative luminosity distribution with 
two power laws at luminosities below and above $5 \times 10^{37}$ ergs s$^{-1}$. For 
source luminosities between $10^{36}$ and $5 \times 10^{37}$ ergs s$^{-1}$, we obtain 
the slope of $-0.3\pm0.1$ with normalization of $38$ sources at $10^{36}$ ergs s$^{-1}$, 
for higher luminosities, the integral slope is $-1.2^{+0.4}_{-0.7}$ (shown as dotted 
lines in Fig. \ref{GCS_XLF}). We also fitted the unbinned differential luminosity with 
a power law with exponential cut-off: $L^{-\alpha}$exp$(-L/L_{\rm cut})$. The resulting 
values of $\alpha$ and the cut-off luminosity are $1.0$ and 
$2.7^{+4.3}_{-0.9} \times 10^{38}$ ergs s$^{-1}$ respectively. There is some flattening 
of the cumulative distribution toward the faint luminosity end, probably caused by 
incompleteness of our sample. 

As it is clearly seen from Fig. \ref{GCS_XLF}, the integrated luminosity function of 
GC sources differs significantly from that of the other point sources in the central bulge 
of M31. The low luminosity part of the GC luminosity distribution is somewhat flatter than 
that of the central bulge of M31. The luminosity of the break in the slope of the luminosity 
function appears to be $\sim 2 - 3$ times higher for GC sources 
($L_{\rm br}^{\rm GC} \sim 5 \times 10^{37}$ ergs s$^{-1}$) than for the sources 
in the central bulge with $L_{\rm br}^{\rm bulge} \sim (1.5 - 2.0) \times 10^{37}$ ergs s$^{-1}$ 
(\cite{Shirey01,lum_distr02,Kong02}). Qualitatively, the fraction of the bright sources 
($L_{\rm X} > few \times 10^{37}$ ergs s$^{-1}$) is higher for the GC population than for 
the central bulge population. It is interesting to note that in spite of the aforementioned 
differences, the slopes of the high-luminosity parts of the luminosity distributions for 
the M31 GC and central bulge populations are very close to each other. 

In addition, there is an obvious difference between the luminosity functions of GCs and X-ray 
sources in the disk of M31 (\cite{lum_distr02}). The luminosity distribution of the disk 
X-ray source population terminates at much lower luminosities ({\em few} $\times 10^{37}$ ergs 
s$^{-1}$) (\cite{lum_distr02}) than that of the M31 GCs ($several \times 10^{38}$ ergs s$^{-1}$).    

It is also interesting to compare the luminosity functions of the Galactic GC and M31 GC 
X-ray sources. These two functions have two main differences: i) the peak X-ray luminosity 
of the M31 GC X-ray sources is $\sim 10$ times higher than that of the brightest Galactic 
GC X-ray source (\cite{DiStefano01}); ii) the fraction of the bright sources is much higher 
for the M31 GC population than for the Galactic GC population; 13 sources or $\sim 30 \%$ of the 
total number of GC sources detected in our survey reached luminosity of $5 \times 10^{37}$ ergs 
s$^{-1}$, while only one Galactic GC source, X1820-303 (NGC 6624) has luminosity occasionally 
exceeding this level (\cite{Sidoli01}).

\section{On the role of the bright GC sources in the X-ray source population of M31.}

We found that $\sim 10 \%$ of the optically identified GC candidates in the regions of M31, covered 
by {\em XMM-Newton} and {\em Chandra} observations, harbor bright X-ray sources with luminosities 
above $10^{35}$ ergs s$^{-1}$ (43 X-ray sources were found among $\sim 400$ GC candidates). This 
fraction is comparable to the observed for the Milky Way globular cluster system with 14 bright X-ray 
sources among known 150 globular clusters (\cite{V95,Sidoli01,Harris96}). In addition, the fraction 
of M31 GCs that contain X-ray sources with luminosities above $\sim 10^{37}$ ergs s$^{-1}$ 
($\sim 5 \%$) appears to be consistent with that derived for a large sample of galaxies of various 
morphological types (\cite{Sarazin03}).

Our observations clearly indicate that GC X-ray sources make dominant contribution to the bright 
($L_{X} > few \times 10^{37}$ ergs s$^{-1}$) source counts in the surveyed areas of M31. Six 
persistent globular cluster X-ray sources (sources $\#\# 2, 9, 22, 41, 42, 43$) have absorbed 
luminosities in the $0.3 - 10$ keV energy band exceeding $10^{38}$ ergs s$^{-1}$; three other 
persistent GC sources in our sample (sources $\#\# 31, 32, 37$) occasionally exceed that luminosity 
level, i.e. at least six GC X-ray sources are brighter than $10^{38}$ ergs s$^{-1}$ at any given 
time. Apart from transient X-ray sources, we detected only one persistent non-GC source belonging 
to M31, having average luminosity in excess of $10^{38}$ ergs s$^{-1}$, RX J0042.6+4115 
(\cite{Barnard1}), and two other non-GC sources occasionally exceeding that luminosity level in 
the same regions. Therefore, from $\sim 67\%$ to $\sim 90 \%$ of the sources brighter than $10^{38}$ 
ergs s$^{-1}$ observed in the surveyed area at any given time are associated with globular clusters. 
According to the results of our survey, about 20 sources or $\sim 40\%$ of the total number of sources 
with luminosities above $10^{37}$ ergs s$^{-1}$ detected in {\em XMM} and {\em Chandra} observations 
of M31, reside in globular clusters. 

We found that fraction of bright X-ray sources ($L_{X} > 10^{37}$ ergs s$^{-1}$), residing in globular 
clusters is significantly higher in M31 ($\sim 40 \%$) than in Milky Way Galaxy ($\sim 5 - 10 \%$)
(\cite{Liu01}). It should be also noted, that no persistent GC sources with luminosities in excess of 
$10^{38}$ ergs s$^{-1}$ have been detected in our Galaxy so far, compared to six sources found in M31. 
On the other hand, the fraction of bright GC X-ray sources in M31 is surprisingly similar to the 
observed in early-type galaxies: i.e. NGC 1399 (\cite{Angelini01}) and NGC 4472 (\cite{Kundu02}).

\section{Spatial distribution and luminosity of M31 GC X-ray sources.}

\begin{figure}
\epsfxsize=18cm
\epsffile{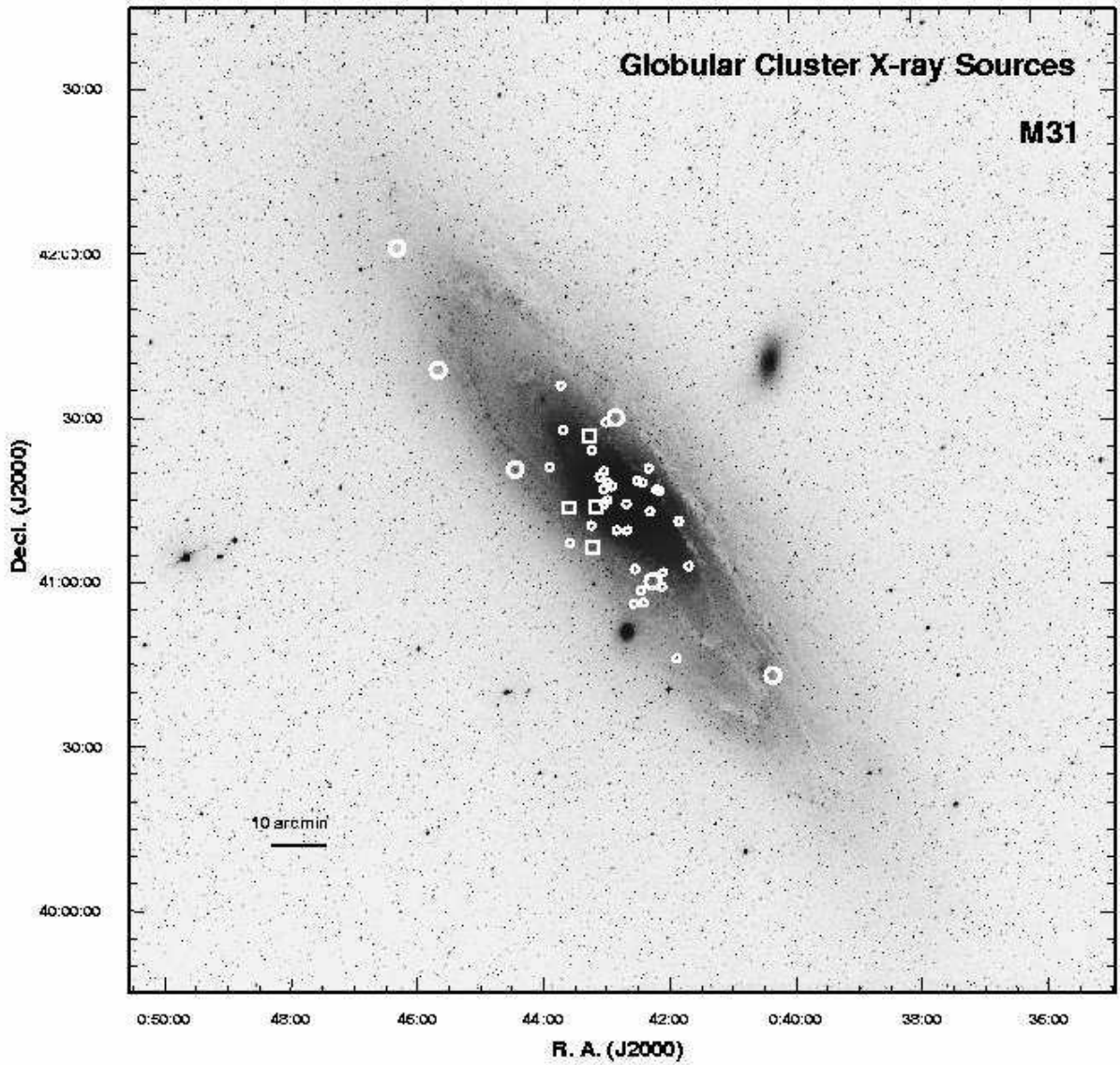}
\caption{Spatial distribution of M31 GC X-ray sources detected with {\em XMM-Newton} and 
{\em Chandra}. The brightest sources with luminosities above $10^{38}$ ergs s$^{-1}$ are 
marked with large white circles. X-ray sources with luminosities occasionally exceeding 
$10^{38}$ ergs s$^{-1}$ are shown with white boxes. The positions of fainter GC X-ray 
sources are shown with small white circles. \label{GCS_spatial_distr}}
\end{figure}

The spatial distribution of GC X-ray sources detected with {\em XMM-Netwon} and {\em Chandra} is 
shown in Fig. \ref{GCS_spatial_distr}. We have noticed that bright X-ray sources tend to reside at 
significantly larger galactocentric distances than the fainter sources. In fact, all GC X-ray sources 
with luminosities steadily exceeding $10^{38}$ ergs  s$^{-1}$ (shown in Fig. \ref{GCS_spatial_distr} 
with {\em large white circles}) happen to have angular distances more than $15 \arcmin$ from the center 
of the galaxy, or well outside the central bulge of M31, where the most of fainter GC X-ray sources are 
residing.  

\section{On the correlation between the X-ray and optical properties.}

We used the results of optical observations of M31 GC candidates (\cite{Huchra91}; 
\cite{Barmby00}; \cite{Perrett02}) to study how the properties of the X-ray sources depend 
on the optical properties of the globular clusters hosting them.  

\begin{figure}
\hbox{
\begin{minipage}{9.0cm}
\epsfxsize=9.0cm
\epsffile{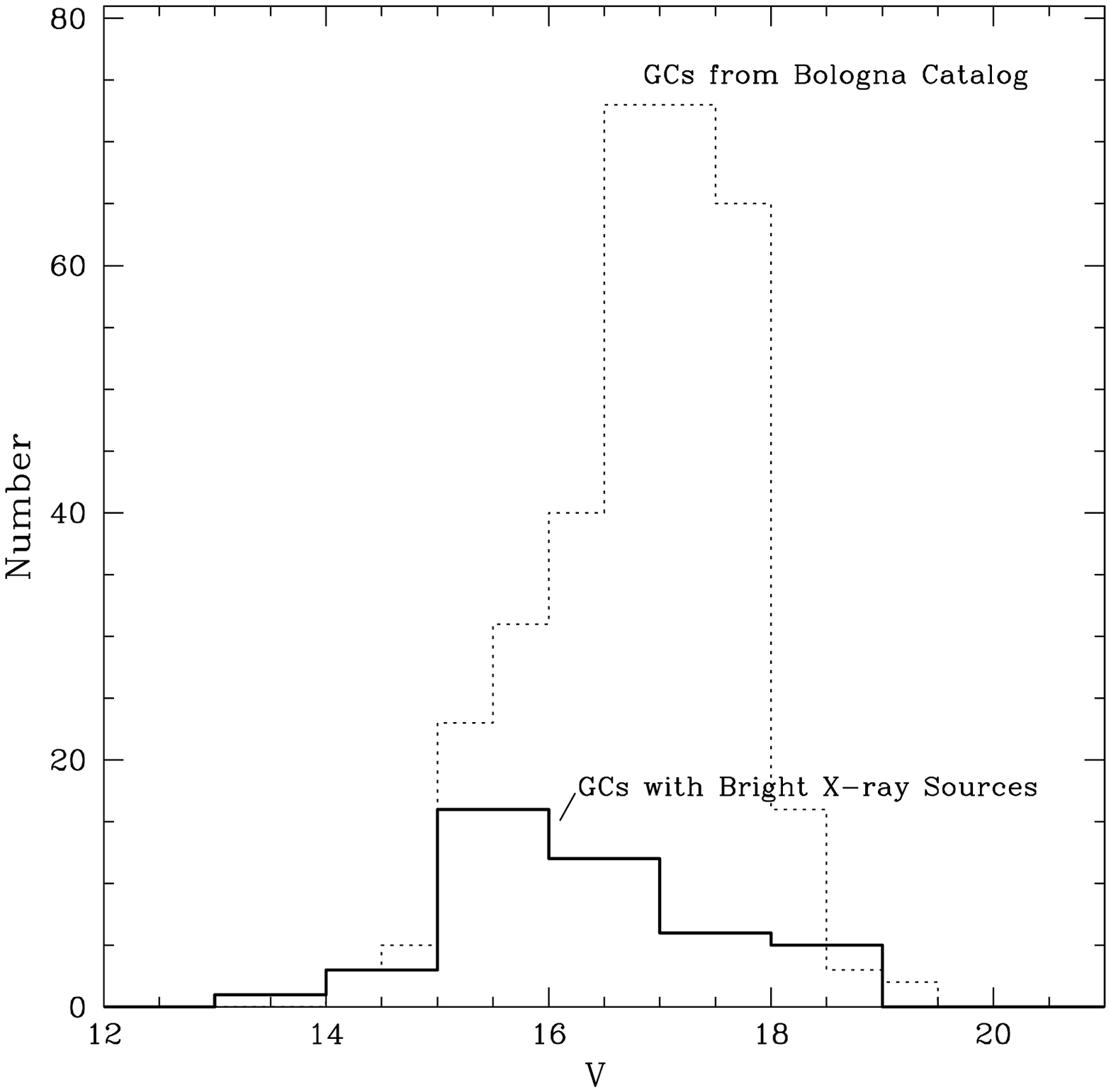}
\end{minipage}
\begin{minipage}{9.0cm}
\epsfxsize=9.0cm
\epsffile{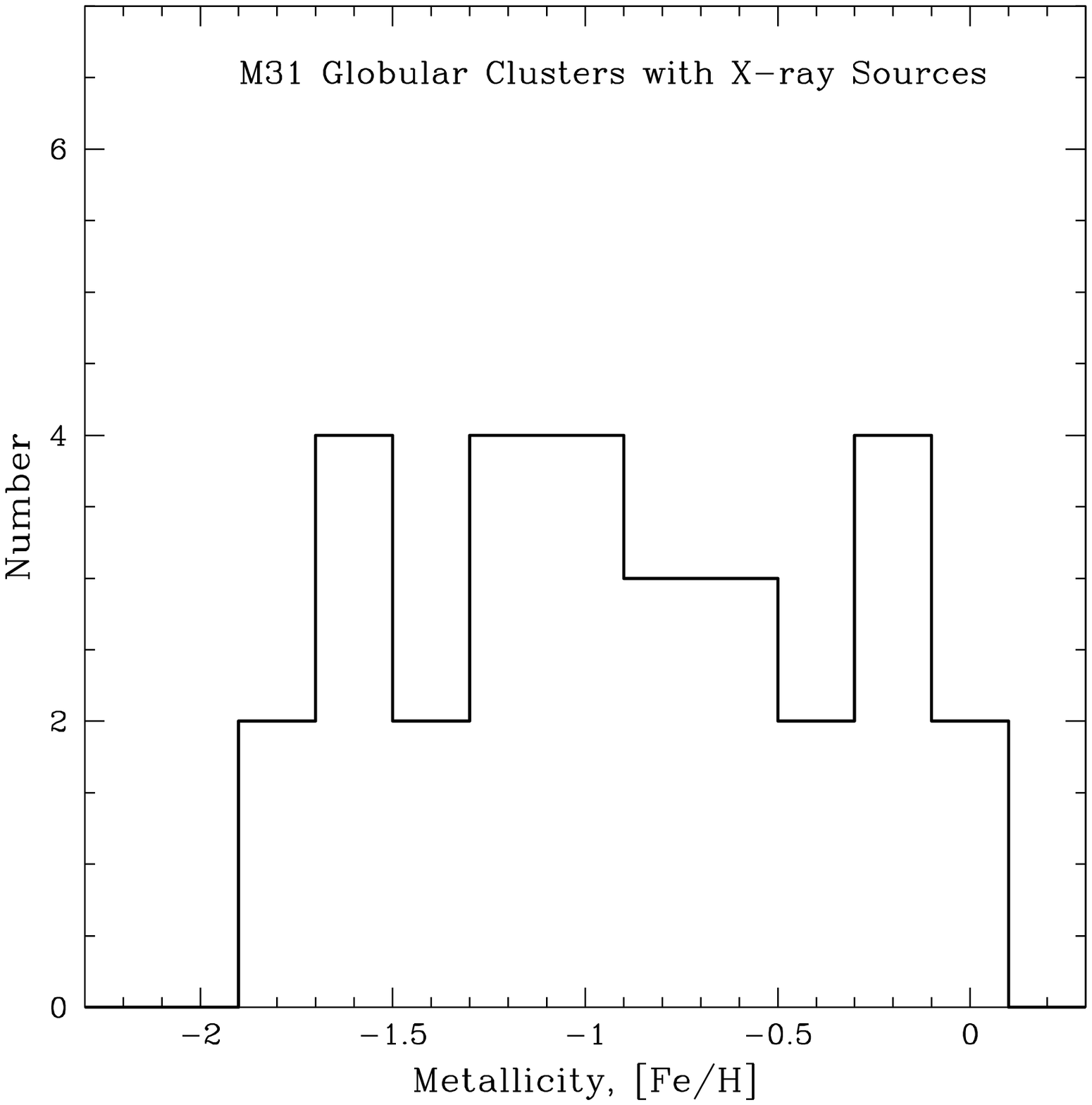}
\end{minipage}
}
\caption{{\em Left panel:} V-magnitude distributions for M31 GC candidates. The results 
for GC hosting bright X-ray sources are shown with {\em thick} histogram. The distribution 
for the GC candidate sample from Battistini et al. 1987 is shown with {\em dotted} histogram. 
{\em Right panel:} metallicity ([Fe/H]) distribution for M31 GC candidates hosting bright 
X-ray sources. \label{GC_opt_distr}}
\end{figure}

We found that globular clusters hosting bright X-ray sources tend to be optically brighter than 
the rest of M31 globular clusters, in agreement with previous studies (\cite{DiStefano01}). 
To illustrate this effect, V-magnitude distribution of X-ray bright globular clusters 
from our sample is shown in Fig. \ref{GC_opt_distr}({\em left panel}) along with the distribution 
for the whole sample of GC candidates from Battistini et al. (1987). Similar tendency for bright 
X-ray sources to be associated with more optically luminous GCs has been also observed in several 
early-type galaxies (\cite{Sarazin03}). Sarazin et al. (2003) shows that this tendency could be 
consistent with a constant probability of finding an LMXB per unit optical luminosity of the cluster. 

Both Galactic and M31 globular clusters hosting bright LMXBs were found to be both denser and 
more metal-rich (\cite{Bellazzini95}). We studied the effect of globular cluster metallicity, 
[Fe/H] on the properties of the M31 GC X-ray sources from our sample. For 31 GC X-ray sources 
the results of metallicity measurements are available (\cite{Huchra91}; \cite{Barmby00}; 
\cite{Perrett02}). The metallicity distribution of the X-ray bright GC in our sample is shown 
in Fig. \ref{GC_opt_distr}({\em right panel}). The population of X-ray emitting GC yields a weighted 
mean $<{\rm [Fe/H]}>_{\rm XR} = -1.06\pm0.02$, significantly higher than  
$<{\rm [Fe/H]}>_{\rm nXR} = -1.37\pm0.01$ derived for the non-X-ray GC subsample from Perrett 
et al. (2002), in agreement with results of previous studies (\cite{Bellazzini95}). We used 
a Kolmogorov-Smirnov (KS) test to determine whether X-ray emitting GC and non-X-ray GC populations 
were drawn from the same distribution. The hypothesis that the two distributions are extracted 
from the same parent population can be rejected at the confidence level of $\sim 93\%$.
 
\begin{figure}
\hbox{
\begin{minipage}{9.0cm}
\epsfxsize=9.0cm
\epsffile{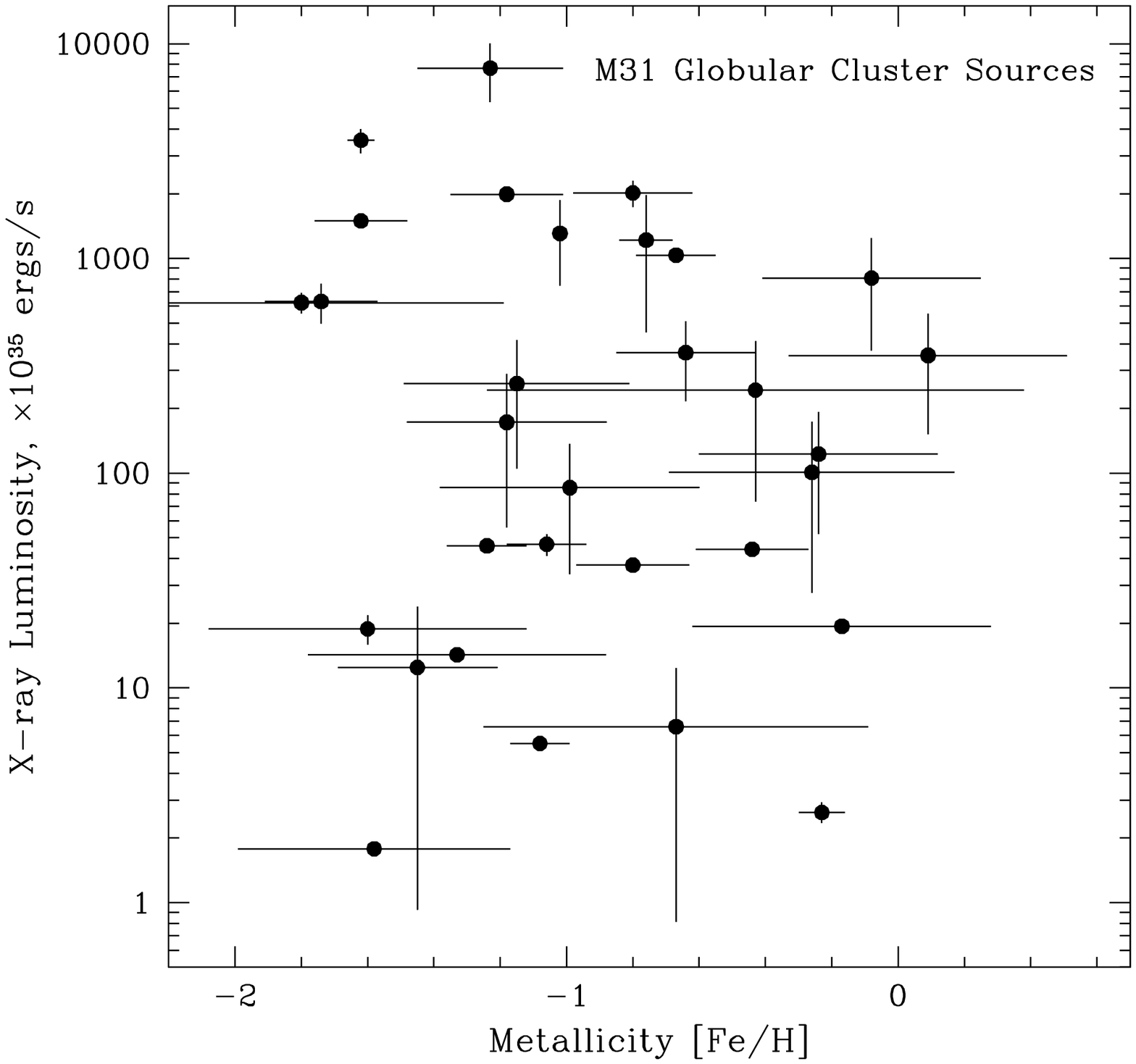}
\end{minipage}
\begin{minipage}{9.0cm}
\epsfxsize=9.0cm
\epsffile{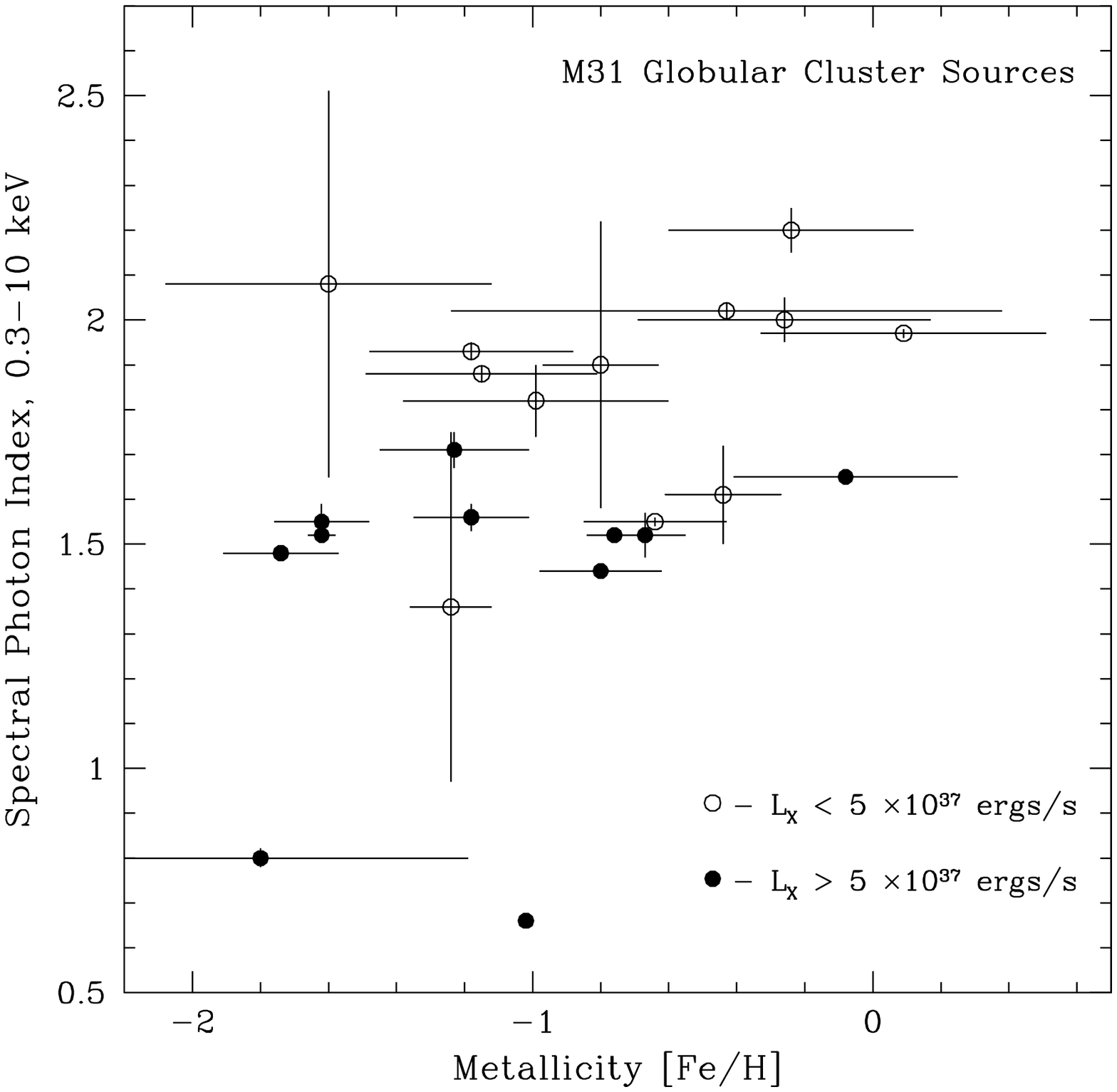}
\end{minipage}
}
\caption{{\em Left panel:} the measured isotropic $0.3 - 10$ keV X-ray luminosity of M31 GC 
X-ray sources versus the host globular cluster metallicity ([Fe/H]). The error bars in Y-axis 
reflect statistical uncertainty of the source flux determination and in some cases the range 
of source X-ray luminosities observed with {\em XMM} and {\em Chandra}. {\em Right panel:} the 
spectral power law photon index for the GC sources from our sample vs. metallicity of the 
globular clusters hosting them. \label{x_ray_prop_metallicity}}
\end{figure}

The dependence of the X-ray luminosity of M31 GC sources on the metallicity of their host 
globular clusters is shown in the {\em left panel} of Fig. \ref{x_ray_prop_metallicity}. The data 
points can be separated into two main groups corresponding to the source X-ray luminosities below 
and above $\sim 10^{38}$ ergs s$^{-1}$. The first group (sources with X-ray luminosities below 
$10^{38}$ ergs s$^{-1}$) has weighted mean metallicity of $-0.79\pm0.04$ and possibly demonstrates 
general tendency of the brighter sources to reside in the more metal rich clusters. This behavior 
could be consistent with possible correlation between X-ray luminosity and cluster metallicity found 
by Sidoli et al. (2001) for the sample of bright Galactic GC sources. It is also interesting to note, 
that all known Galactic GC X-ray sources have luminosities in the range occupied by the first group 
(\cite{Sidoli01}). The whole second group (sources with X-ray luminosities above $10^{38}$ ergs 
s$^{-1}$) occupies low metallicity region with weighted mean metallicity of $-1.12\pm0.02$.

Recent X-ray observations and theoretical work suggest that spectral hardness of the bright LMXB 
sources in the soft X-ray band ($0.1 - 2.4$ keV) can depend on the average metallicity of the globular 
clusters harboring them. The soft X-ray spectra of the sources located in the metal-poor GCs were 
found to be harder than those of the metal-rich clusters (\cite{IB99}; \cite{MKZ04}). To demonstrate 
the effect of cluster metallicity on the intrinsic spectral hardness of the M31 GC X-ray sources, 
in the {\em right panel} of Fig. \ref{x_ray_prop_metallicity} the spectral power law photon index in 
the $0.3 - 7$ keV energy band for the GC sources from our sample is shown as a function of metallicity 
of the globular clusters hosting them. As it is seen from Fig. \ref{x_ray_prop_metallicity} 
({\em right panel}), there is a possible correlation of the X-ray spectral hardness and the metallicity. 
We note however, that it is premature to draw any conclusions on the relationship between the intrinsic 
spectral parameters of X-ray sources and metallicity of their host clusters, based on a model-dependent 
presentation of the spectrum measured in the narrow energy band. For example, using a multi-component 
spectral modeling of the {\em BeppoSAX} broad-band data for the bright Galactic GC sources, Sidoli et al. 
(2001) found a correlation between the temperature of the soft spectral component and a cluster metallicity, 
opposite to the results obtained in the soft X-ray band with one-component spectral models (\cite{IB99}).

\section{On the nature of M31 GC X-ray sources}

Several GC X-ray sources observed in M31 have luminosities exceeding $10^{38}$ ergs s$^{-1}$, being 
much brighter than any of the Galactic GC X-ray sources (Table \ref{source_ID}). Four X-ray sources 
in our sample ($\#\# 2, 10, 22, 42$) have luminosities close to or exceeding a formal Eddington limit 
for the $1.4 M_{\odot}$ neutron star. Nevertheless, such a high luminosities are not uncommon for bright 
low-mass X-ray binaries containing neutron star primaries. The Z-source subclass of Galactic LMXB is 
characterized by high X-ray luminosities ($> 10^{38}$ ergs s$^{-1}$) (\cite{HK89}) with two sources, 
Sco X-1 and GX 5-1, regularly exceeding Eddington limit for a neutron star (\cite{Liu01}). Recently, 
Barnard, Kolb and Osborne (2003) found that bright M31 X-ray source, RX J0042.6+4115 with extremely 
high X-ray luminosity of $\sim 5 \times 10^{38}$ ergs s$^{-1}$ could be also classified as a neutron 
star Z-source. On the other hand, among stellar mass black hole X-ray sources, only transients in 
outburst (\cite{Chen97}) and highly variable systems like GRS 1915+105 (\cite{Morgan97}) reach 
luminosities observed for bright M31 GC sources.   

All bright X-ray sources found in the Milky Way globular clusters show Type I X-ray bursts and 
probably contain neutron star primaries. Moreover, theoretical calculations predict the probability 
of finding a black hole X-ray binary in the globular cluster to be much smaller than the neutron star 
binary due to dynamical effects and extremely low duty cycles of black hole binaries 
(\cite{Kulkarni93,SH93,KKR04}). 
The considerations mentioned above, and a general similarity between the M31 and Milky Way globular 
cluster systems allow us to expect most of the persistent M31 GC X-ray sources to be LMXB systems 
containing a neutron star primaries. However, the possibility of finding an active accreting black 
holes in our GC source sample can not be ruled out, based on the current spectral and timing data. 
For example, the X-ray source RX J0043.2+4127 associated with globular cluster Bo 163 shows a pattern 
of recurrent transient outbursts and quiescent periods (Fig. \ref{GCS_N_CH_003_lc}) typical for both 
neutron star and black hole transient systems. The study of short-term variability of GC X-ray sources 
in M31 with future missions can provide a definitive answer, if Type I X-ray bursts or X-ray pulsations 
are observed from these objects. The broad-band long-term monitoring of the transient/highly variable 
GC sources could also provide valuable information on their spectral evolution helping to establish the 
nature of these systems.  

Some of the M31 GC X-ray sources in our sample might be unresolved composites of separate sources 
(\cite{Angelini01}; \cite{DiStefano01}). Recent high-resolution {\em Chandra} observations of the 
Galactic globular clusters revealed that one of the bright Galactic GC sources is a composite of 
two X-ray sources of comparable brightness (\cite{WA01}). Depending on the probability of finding 
a bright X-ray source in a GC, we can expect to find $2 - 4$ GC X-ray sources in our sample (or 
$\sim 5 - 10 \%$ of the total number) to be composites of two independent sources (\cite{DiStefano01}).

\section{Summary}

Utilizing the data of {\em XMM-Newton} and {\em Chandra} observations, we performed a study of individual 
and group properties of 43 X-ray sources identified with globular clusters in M31. The correlation between 
the optical parameters of the host globular clusters and the properties of X-ray sources associated with 
them was also studied. We compare the individual and group properties of M31 GC X-ray sources with the 
properties of their Galactic counterparts.

The spectral properties of the bright GC X-ray sources in our sample were found to be similar to that 
of the LMXB located in the bulge and globular clusters of the Milky Way Galaxy at the same luminosity 
levels (\cite{WSP88}; \cite{CS97}; \cite{CBC01}). We note, that some GC sources could be unresolved 
composites of two or even more sources of comparable brightness complicating direct comparison with 
their Galactic counterparts. For the majority of sources, we obtained acceptable fits using a power law 
spectral model approximation. However, for several sources with high luminosities, the models with 
quasi-exponential cut-off at $\sim 2 - 8$ keV or two-component models describe the energy spectra 
significantly better than a simple power law. Several sources demonstrate a correlation between the level 
of X-ray flux and hardness of their energy spectrum similar to the Galactic Z and atoll-sources 
(\cite{HK89}). The spectral distribution of M31 GC X-ray sources resembles the corresponding distributions 
derived for the the central bulge of M31 (\cite{Shirey01}) and other nearby galaxies of different 
morphological type (\cite{Irwin03}). 

There is no evidence that X-ray variability of M31 GC sources and bright Galactic LMXB (both GC and non-GC) 
differ. We found that $\sim 80\%$ of the sources in our sample with multiple flux measurements available 
show significant variability on a time scales from days to years. We note, that in some GC X-ray sources 
the observed variability could be result of superposition of flux variations of several unresolved sources. 
The X-ray source RX J0043.2+4127 in Bo 163 has been found to show recurrent transient outbursts with peak 
luminosities above $10^{38}$ ergs s$^{-1}$, typical for Galactic recurrent transients that harbor both 
neutron stars and black holes. Several sources show significant variability on a time scale of individual 
observations, ranging from aperiodic fluctuations to regular dipping (Bo 158).

The X-ray luminosity function of GC sources is found to be significantly different from that of the other 
point sources in the bulge and disk of M31 and Galactic GC X-ray sources. The luminosity distributions of 
the GC X-ray sources in M31 and that of the Milky Way have two main differences: i) the peak luminosity of 
the M31 GC X-ray sources is $\sim 10$ times higher than that of the brightest Galactic GC X-ray source; ii) 
the fraction of the bright sources is much higher for the M31 GC population; $\sim 30 \%$ of the GC sources 
detected in our survey reached the luminosity of $5 \times 10^{37}$ ergs s$^{-1}$, while only one of 14 
Galactic GC sources, X1820-303 (NGC 6624) has luminosity occasionally exceeding this level (\cite{Sidoli01}).

We found that $\sim 10 \%$ of the optically identified GC candidates in the regions of M31, covered by 
{\em XMM-Newton} and {\em Chandra} observations, harbor bright X-ray sources with luminosities above 
$10^{35}$ ergs s$^{-1}$. This fraction is comparable to that observed for the Milky Way globular cluster 
system with 14 bright X-ray sources among known $\sim 150$ globular clusters (\cite{V95,Sidoli01,Harris96}). 
In addition, the fraction of M31 GCs that contain X-ray sources with luminosities above $\sim 10^{37}$ 
ergs s$^{-1}$ ($\sim 5 \%$) appears to be consistent with that derived for a large sample of galaxies of 
different morphological type (\cite{Sarazin03}).    

The inferred isotropic X-ray luminosities of the GC sources lie between $\sim 10^{35}$ and $\sim 10^{39}$ 
erg s$^{-1}$ in the $0.3 - 10$ keV energy band. Six persistent sources (or $\sim 14 \%$ of the total 
number) have luminosities steadily exceeding $10^{38}$ erg s$^{-1}$, and three other sources occasionally 
exceed that luminosity level. Four X-ray sources in our sample ($\#\# 2, 10, 22, 42$) have luminosities 
close to or exceeding the Eddington limit for the $1.4 M_{\odot}$ neutron star. Our observations indicate 
that GC sources make dominant contribution to the bright source counts in the areas of M31 covered by 
the survey: $\sim 40 \%$ of the total number of sources with luminosities above $10^{37}$ ergs s$^{-1}$ 
reside in GCs. The fraction of the GC X-ray sources rises to $67 - 90\%$ if one considers source 
luminosities above $10^{38}$ erg s$^{-1}$. The contribution of the GC sources to the total number of 
bright sources found in M31 is much higher (by factor of $>4$) than in the Milky Way galaxy. On the other 
hand, there is a surprising similarity between the fractions of bright GC X-ray sources in M31 and in the 
early-type galaxies (\cite{Angelini01,Kundu02}). 

The distribution of radial distances of GC X-ray sources from the center of M31 shows that brightest 
systems with luminosities steadily exceeding $10^{38}$ erg s$^{-1}$ reside at greater galactocentric 
distances than the rest of the GC sources.

We used the results of optical observations of M31 GC candidates to study how the properties of the 
X-ray sources depend on the optical properties of the globular clusters hosting them. We found that 
globular clusters hosting bright X-ray sources tend to be optically brighter and more metal rich than 
the rest of M31 globular clusters, in agreement with previous studies (\cite{Bellazzini95,DiStefano01}). 
The comparison of the X-ray properties of GC sources with metallicity of their host globular clusters 
has shown that the brightest sources with luminosities above $\sim 10^{38}$ ergs s$^{-1}$ tend to reside 
in more metal poor clusters than the fainter sources. 

The remarkable similarities between the properties of the M31 GC X-ray sources and that of the Galactic 
neutron star LMXBs allow us expect most of the persistent M31 GC X-ray sources to be LMXB systems with 
neutron star primaries. We note, however, that current X-ray spectral and timing data can not rule out 
the possibility of finding an active accreting black holes in our GC source sample. The study of 
short-term variability and broad-band spectral evolution of GC X-ray sources in M31 with future 
missions can provide valuable information allowing to establish the nature of these systems.

In conclusion, we note that in spite of the similarities of the spectral shape and variability patterns 
of individual M31 GC X-ray sources and Galactic LMXBs (both GC and non-GC), the group properties of M31 
and Milky Way globular cluster X-ray sources differ significantly. The M31 GC source population has a 
much higher fraction of bright X-ray sources and higher peak luminosity than the Milky Way GC population, 
probably suggesting different formation histories and evolution. The large fraction of GC systems among 
the bright X-ray sources in M31 makes the M31 population of GC X-ray sources more like giant early-type 
galaxies than the Milky Way.

\section{Acknowledgments}
Part of this work was done during a summer workshop at the Aspen Center for Physics, authors are grateful 
to the Center for their hospitality. Support for this work was provided through NASA Grant NAG5-12390. 
This research has made use of data obtained through the {\em Chandra} public data archive. {\em XMM-Newton} 
is an ESA Science Mission with instruments and contributions directly funded by ESA Member states and the 
USA (NASA). This research has made use of data obtained through the High Energy Astrophysics Science Archive 
Research Center Online Service, provided by the NASA/Goddard Space Flight Center.

\begin{table}
\small
\caption{{\em XMM-Newton} observations of M31 fields used in our analysis. 
\label{obslog_xmm}}
\begin{tabular}{cccccccc}
\hline
\hline
Obs. $\#$&Date, UT&Field&Obs. ID&RA (J2000)$^{a}$&Dec (J2000)$^{a}$&Exp.(MOS)$^{b}$&Exp.(pn)$^{b}$\\
         &        &     &       &  (h:m:s)       &(d:m:s)          &(ks)           &(ks)\\             
\hline
 $\#1$&2000 Jun 25 &M31 Core  &0112570401&00:42:43.0&41:15:46.1&28.9&24.9\\
 $\#2$&2000 Dec 28 &M31 Core  &0112570601&00:42:43.0&41:15:46.1&12.1& 9.4\\
 $\#3$&2001 Jan 11 &M31 G1    &0065770101&00:32:46.9&39:34:41.7& 7.3& 4.9\\
 $\#4$&2001 Jun 29 &M31 Core  &0109270101&00:42:43.0&41:15:46.1&29.0&24.9\\
 $\#5$&2002 Jan 05 &M31 North1&0109270701&00:44:01.0&41:35:57.0&57.3&54.7\\
 $\#6$&2002 Jan 06 &M31 Core  &0112570101&00:42:43.0&41:15:46.1&63.0&49.9\\
 $\#7$&2002 Jan 13 &M31 South1&0112570201&00:41:25.0&40:55:35.0&53.9&53.3\\
 $\#8$&2002 Jan 24 &M31 South2&0112570301&00:40:06.0&40:35:24.0&30.0&29.0\\
 $\#9$&2002 Jan 26 &M31 North2&0109270301&00:45:20.0&41:56:09.0&29.1&25.3\\
$\#10$&2002 Jun 29 &M31 North3&0109270401&00:46:38.0&42:16:20.0&51.5&36.5\\
\hline
\end{tabular}
\begin{list}{}{}
\item[$^{a}$] -- coordinates of the center of the field of view
\item[$^{b}$] -- instrument exposure used in the analysis 
\end{list}
\end{table}

\clearpage

\begin{table}
\small
\caption{{\em Chandra} observations of M31 fields used in our analysis. 
\label{obslog_chandra}}
\begin{tabular}{ccccccc}
\hline
\hline
Obs. $\#$ & Date,UT     & Field & Obs. ID/Instrument  & R.A. (J2000)$^{a}$ & Decl. (J2000)$^{a}$ & Exp.$^{b}$\\
          & (yyyy-mo-dd)&       &          &  (h:m:s)         &(d:m:s)            &   (ks)\\             
\hline
&1999-10-13&M31 NUCLEUS   & 303/ACIS-I&00:42:42.59&41:16:12.2& 12.0\\
&1999-12-11&M31 TRANSIENT & 305/ACIS-I&00:42:44.40&41:16:08.3&  4.1\\
&1999-12-23&M31 CENTER    & 268/HRC-I &00:42:44.40&41:16:08.3&  5.2\\
&1999-12-27&M31 TRANSIENT & 306/ACIS-I&00:42:44.40&41:16:08.3&  4.1\\
&2000-01-29&M31 TRANSIENT & 307/ACIS-I&00:42:44.40&41:16:08.3&  4.1\\
&2000-02-16&M31 TRANSIENT & 308/ACIS-I&00:42:44.40&41:16:08.3&  4.0\\
&2000-03-08&M31 CENTER    & 271/HRC-I &00 42 44.40&41 16 08.3&  2.4\\
&2000-06-01&M31 TRANSIENT & 309/ACIS-S&00:42:44.40&41:16:08.3&  5.1\\
&2000-07-02&M31 TRANSIENT & 310/ACIS-S&00:42:44.40&41:16:08.3&  5.1\\
&2000-07-29&M31 TRANSIENT & 311/ACIS-I&00:42:44.40&41:16:08.3&  4.9\\
&2000-08-27&M31 TRANSIENT & 312/ACIS-I&00:42:44.40&41:16:08.3&  4.7\\
&2000-09-21&M31 TRANSIENT & 313/ACIS-S&00:42:40.80&40:51:54.0&  6.1\\
&2000-10-21&M31 TRANSIENT & 314/ACIS-S&00:42:40.80&40:51:54.0&  5.1\\
&2000-11-01&M31-3         &2052/ACIS-S&00:46:16.70&41:40:55.0& 14.1\\
&2000-11-05&M31-2         &2049/ACIS-S&00:41:49.90&40:59:20.0& 14.8\\
&2000-11-17&M31 TRANSIENT &1580/ACIS-S&00:42:40.80&40:51:54.0&  5.1\\
&2000-12-13&M31 NUCLEUS   &1581/ACIS-I&00:42:44.40&41:16:08.3&  4.4\\
&2001-01-13&M31 TRANSIENT &1854/ACIS-S&00:42:40.80&41:15:54.0&  4.7\\
&2001-02-18&M31 NUCLEUS   &1582/ACIS-I&00:42:44.40&41:16:08.3&  4.3\\
&2001-03-08&M31-3         &2053/ACIS-S&00:46:16.70&41:40:55.0& 13.5\\
&2001-03-08&M31-2         &2050/ACIS-S&00:41:49.90&40:59:20.0& 13.2\\
&2001-06-10&M31 NUCLEUS   &1583/ACIS-I&00:42:44.40&41:16:08.3&  5.0\\
&2001-07-03&M31 NUCLEUS   &1584/ACIS-I&00:42:37.50&40:54:27.0&  4.9\\
&2001-07-03&M31-3         &2054/ACIS-S&00:46:16.70&41:40:55.0& 14.7\\
&2001-07-03&M31-2         &2051/ACIS-S&00:41:49.90&40:59:20.0& 13.8\\
&2001-07-24&M32           &2017/ACIS-S&00:42:41.80&40:51:52.0& 46.4\\
&2001-07-28&M32           &2494/ACIS-S&00:42:41.80&40:51:52.0& 16.1\\
&2001-08-31&M31 NUCLEUS   &1577/ACIS-I&00:43:08.50&41:18:20.0&  4.9\\
&2001-10-05&M31 NUCLEUS   &1575/ACIS-S&00:42:44.40&41:16:08.3& 38.1\\
&2001-10-05&M31 NUCLEUS   &1576/ACIS-I&00:42:34.89&40:57:21.0&  4.9\\
&2001-10-31&M31           &1912/HRC-I &00:42:42.30&41:16:08.4& 46.9\\
&2001-11-19&M31 NUCLEUS   &1585/ACIS-I&00:43:05.55&41:17:03.3&  4.9\\
&2001-12-07&M31 TRANSIENT &2895/ACIS-I&00:43:08.50&41:18:20.0&  4.9\\
&2002-01-08&M31 TRANSIENT &2897/ACIS-I&00:43:09.80&41:19:00.7&  4.9\\
&2002-02-06&M31 TRANSIENT &2896/ACIS-I&00:43:05.50&41:17:03.3&  4.9\\
&2002-06-02&M31 TRANSIENT &2898/ACIS-I&00:43:09.80&41:19:00.7&  4.9\\
&2002-07-08&M31 TRANSIENT &2901/ACIS-I&00:41:54.67&40:56:47.5&  4.6\\
&2002-08-11&M31           &4360/ACIS-I&00:42:44.40&41:16:08.9&  4.9\\
&2002-08-23&M31 TRANSIENT &2899/ACIS-I&00:41:54.69&40:56:47.8&  4.9\\
&2002-10-14&M31 TRANSIENT &2894/ACIS-I&00:42:34.90&40:57:21.0&  4.7\\
\hline
\end{tabular}
\begin{list}{}{}
\item[$^{a}$] -- coordinates of the center of the field of view
\item[$^{b}$] -- total exposure 
\end{list}
\end{table}

\begin{table}
\small
\caption{List of GC X-ray sources detected in {\em XMM-Newton} and {\em Chandra} 
observations of M31. \label{source_ID}}
\begin{tabular}{cccccl}
\hline
\hline
Source& R.A. & Decl. &$L_{\rm X}$(0.3-10 keV)         & Optical ID$^{a}$& X-ray ID$^{b}$\\ 
  ID  &      &       &($\times 10^{35}$ ergs s$^{-1}$)&                 & \\     
\hline
 1& 00 32 46.6&39 34 40& 6          & G1             & \\
 2& 00 40 20.3&40 43 57& 1990       & Bo5            & SHP73 \\
 3& 00 41 41.0&41 04 01& 3          & MIT87          & D27 \\
 4& 00 41 50.5&41 12 12& 2.4-2.9    & Bo55           & \\
 5& 00 41 52.9&40 47 10& 1-24       & Bo58,MIT106    & D22 \\
 6& 00 42 06.1&41 02 48& 31-73      & Bo D42,MIT130  & SHP138,D13 \\
 7& 00 42 07.1&41 00 16& 40-100     & Bo D44         & D15 \\
 8& 00 42 09.5&41 17 45& 40-114     & MIT140         & D10 \\
 9& 00 42 12.1&41 17 58& 24-300     & Bo78,MIT153    & D20 \\
10& 00 42 15.8&41 01 14& 1735-2330  & Bo82,MIT159    & SHP150,D2 \\
11& 00 42 18.6&41 14 01& 498-763    & Bo86,MIT164    & SHP158,D4 \\
12& 00 42 19.6&41 21 53& 5          & MIT165/MIT166  & D24 \\
13& 00 42 25.0&40 57 19& 19         & Bo94,MIT173    & SHP168,D21 \\
14& 00 42 26.1&41 19 15& 28-174     & Bo96,MIT174    & D14 \\
15& 00 42 27.4&40 59 36& 1-12       & Bo98           & D25 \\
16& 00 42 31.2&41 19 38& 56-290     & Bo107,MIT192   & SHP175,D16 \\
17& 00 42 33.1&41 03 28& 41-52      & Bo110          & SHP178 \\
18& 00 42 34.4&40 57 09& 14         & Bo117          & \\
19& 00 42 40.6&41 10 32& 16-27      & Bo123,MIT212   & D18 \\
20& 00 42 41.4&41 15 23& 34-137     & MIT213         & D23 \\
21& 00 42 50.7&41 10 33& 3          & MIT222         & D28 \\
22& 00 42 51.9&41 31 07& 3093-4009  & Bo135          & SHP205 \\
23& 00 42 55.5&41 18 35& 8-84       & Bo138          & \\
24& 00 42 59.6&41 19 19& 152-555    & Bo143          & SHP217,D5 \\
25& 00 42 59.8&41 16 05& 216-512    & Bo144          & D6 \\
26& 00 43 01.3&41 30 17& 553-692    & Bo91,MIT236    & SHP218 \\
27& 00 43 02.9&41 15 22& 74-414     & Bo146          & SHP220,D7 \\
28& 00 43 03.3&41 21 22& 52-194     & Bo147,MIT240   & SHP222,D11 \\
29& 00 43 03.8&41 18 05& 105-418    & Bo148          & SHP223,D8 \\
30& 00 43 07.5&41 20 20& 16-72      & Bo150,MIT246   & D19 \\
31& 00 43 10.6&41 14 50& 373-1248   & Bo153,MIT251   & SHP228,D3 \\
32& 00 43 14.2&41 07 24& 600-1880   & Bo158          & SHP229,D12 \\
33& 00 43 14.7&41 25 13& 2          & Bo159          & \\
34& 00 43 15.4&41 11 24& 16-22      & Bo161,MIT260   & D26 \\
35& 00 43 17.8&41 27 45& $<1-1010$  & Bo163          & RX J0043.2+4127 \\
36& 00 43 36.7&41 08 10& 46         & Bo182          & \\
37& 00 43 37.3&41 14 43& 454-1981   & Bo185,MIT299   & SHP247,D9 \\
38& 00 43 42.9&41 28 50& 52         & MIT311         & SHP250 \\
39& 00 43 45.5&41 36 57& 44         & Bo193          & SHP253 \\
40& 00 43 56.5&41 22 02& 37         & Bo204          & SHP261 \\
41& 00 44 29.6&41 21 36& 1130       & Bo225          & SHP282 \\
42& 00 45 45.4&41 39 42& 5148-10372 & Bo375          & SHP318,D1 \\
43& 00 46 27.0&42 01 51& 1496       & Bo386          & SHP349 \\
\hline
\end{tabular}

\begin{list}{}
\item $^{a}$ -- source identifications beginning with Bo refer to Globular Cluster 
candidates listed in Table IV of Battistini et al. (1987), MIT -- in Magnier (1993).
\item $^{b}$ -- source identifications beginning with SHP refer to M31 {\em ROSAT}/PSPC 
X-ray source catalog entries from Supper et al. (1997, 2001). Identifications beginning 
with D refer to the list of GC X-ray sources listed in Table 3 of DiStefano et al. (2002). 

\end{list}
\end{table}

\clearpage

\begin{table}
\tiny
\caption{Best-fit model parameters of the energy spectra of the bright GC sources. 
{\em XMM-Newton}/EPIC data, $0.3 - 10$ keV energy range. Absorbed simple power law 
model approximation. Parameter errors correspond to the $1 \sigma$ level. 
\label{spec_par_GCS_xmm_powerlaw}}
\begin{tabular}{ccccccc}
\hline
\hline
ID $^{a}$ & & & & & & Observation/Instrument \\
\hline
              & \multicolumn{5}{c}{Model: Absorbed Power Law (POWERLAW*WABS)}&        \\
\hline
              & Photon   & N$_{\rm H}^{b}$           &Flux$^{c}$&$\chi^{2}$&$L_{X}^{d}$&     \\
              & Index    &                           &          & (dof)    & &    \\
\hline
 2&$1.55\pm0.03$&$0.15\pm0.01$&$28.80\pm0.32$&186.3(181)&$1990$& xmm 8/(MOS1+MOS2)\\
\hline
 6&$1.84^{+0.17}_{-0.10}$&$<0.05$&$0.81\pm0.06$&32.1(29) &$56$  & xmm 7/(MOS1+MOS2)\\
\hline
 7&$1.70^{+0.19}_{-0.16}$&$0.31^{+0.09}_{-0.06}$&$1.44\pm0.07$&50.4(49)&$99$& xmm 7/(MOS1+MOS2)\\
\hline
 8&$2.18^{+0.13}_{-0.15}$&$0.33^{+0.03}_{-0.05}$&$0.93\pm0.07$&30.2(29)&$64$& xmm 1/pn\\
  &$2.33^{+0.25}_{-0.18}$&$0.45^{+0.08}_{-0.07}$&$0.80\pm0.06$&23.8(30)&$55$& xmm 4/pn\\
  &$2.13^{+0.08}_{-0.11}$&$0.37^{+0.02}_{-0.04}$&$1.65\pm0.07$&72.6(73)&$114$& xmm 6/pn\\
\hline
 9&$2.53^{+0.21}_{-0.28}$&$0.61^{+0.20}_{-0.11}$&$0.58\pm0.06$&18.2(15)&$40$& xmm 1/pn\\
  &$2.44^{+0.20}_{-0.26}$&$0.54\pm0.10$&$0.74\pm0.06$&29.7(33)&$51$& xmm 4/pn\\
  &$1.83^{+0.08}_{-0.06}$&$0.51^{+0.06}_{-0.04}$&$3.25\pm0.10$&83.5(105)&$225$& xmm 6/pn\\
\hline
10&$1.37^{+0.05}_{-0.07}$&$0.44\pm0.03$&$32.20\pm0.64$&94.6(94)&$2226$& xmm 4/pn\\
  &$1.52\pm0.03$&$0.49^{+0.01}_{-0.02}$&$31.40\pm0.30$&231.2(293)&$2170$& xmm 7/(MOS1+MOS2+pn)\\
\hline
11&$1.42^{+0.04}_{-0.07}$&$0.09\pm0.02$&$9.72\pm0.20$&115.2(113)&$672$& xmm 1/(MOS1+MOS2)\\
  &$1.34\pm0.06$&$0.09\pm0.02$&$9.45\pm0.29$&72.5(60)&$653$& xmm 2/pn\\
  &$1.61\pm0.04$&$0.14^{+0.01}_{-0.04}$&$8.29\pm0.25$&195.9(189)&$573$& xmm 4/(MOS1+pn)\\
  &$1.56\pm0.03$&$0.12\pm0.01$&$8.27\pm0.09$&563.1(508)&$571$& xmm 6/(MOS1+MOS2+pn)\\
\hline
14&$1.90^{+0.07}_{-0.06}$&$0.25^{+0.05}_{-0.01}$&$1.87\pm0.06$&125.7(97)&$129$& xmm 1/(MOS1+MOS2+pn)\\
  &$1.81\pm0.06$&$0.31^{+0.03}_{-0.02}$&$2.20\pm0.06$&157.7(145)&$152$& xmm 4/(MOS1+MOS2+pn)\\
  &$2.30^{+0.05}_{-0.06}$&$0.31^{+0.04}_{-0.02}$&$0.98\pm0.04$&127.2(138)&$68$& xmm 6/(MOS1+MOS2+pn)\\
\hline
16&$1.84^{+0.08}_{-0.07}$&$0.15^{+0.02}_{-0.03}$&$1.18\pm0.05$&93.0(89)&$82$& xmm 1/(MOS1+MOS2+pn)\\
  &$1.78^{+0.19}_{-0.08}$&$0.20^{+0.07}_{-0.05}$&$2.73\pm0.20$&34.6(26)&$190$& xmm 2/pn\\
  &$1.91^{+0.04}_{-0.05}$&$0.22^{+0.01}_{-0.03}$&$3.36\pm0.07$&184.5(209)&$232$& xmm 4/(MOS1+MOS2+pn)\\
  &$2.17^{+0.04}_{-0.03}$&$0.27^{+0.01}_{-0.04}$&$2.12\pm0.04$&208.8(243)&$147$& xmm 6/(MOS1+MOS2+pn)\\
\hline
19&$1.52^{+0.34}_{-0.19}$&$<0.06$&$3.10\pm0.28$&70.8(62)&$21$& xmm 4/pn+xmm 6/pn\\
\hline
20&$1.82^{+0.07}_{-0.08}$&$0.12\pm0.02$&$1.99\pm0.08$&238.1(214)&$137$& xmm 6/pn\\
\hline
22&$1.57^{+0.03}_{-0.02}$&$0.26\pm0.01$&$52.95\pm0.42$&443.7(424)&$3651$& xmm 5/(MOS1+MOS2)\\
\hline
24&$1.92^{+0.03}_{-0.04}$&$0.13\pm0.01$&$4.39\pm0.09$&213.4(199)&$304$& xmm 1/(MOS1+MOS2+pn)\\
  &$2.11\pm0.11$&$0.18^{+0.02}_{-0.05}$&$4.69\pm0.19$&51.0(63)&$324$& xmm 2/pn\\
  &$1.94^{+0.05}_{-0.04}$&$0.14\pm0.01$&$5.21\pm0.09$&304.0(284)&$360$& xmm 4/(MOS1+MOS2+pn)\\
  &$1.92\pm0.02$&$0.12\pm0.01$&$5.68\pm0.08$&486.3(450)&$393$& xmm 6/(MOS1+MOS2+pn)\\
\hline
25&$1.60\pm0.04$&$0.13\pm0.01$&$5.31\pm0.10$&197.7(201)&$367$& xmm 1/(MOS1+MOS2+pn)\\
  &$1.48^{+0.03}_{-0.01}$&$0.09^{+0.02}_{-0.01}$&$7.64\pm0.11$&320.6(316)&$528$& xmm 4/(MOS1+MOS2+pn)\\
  &$1.57^{+0.02}_{-0.03}$&$0.13\pm0.01$&$7.15\pm0.09$&524.7(517)&$495$& xmm 6/(MOS1+MOS2+pn)\\
\hline
26&$0.83^{+0.04}_{-0.03}$&$0.07^{+0.01}_{-0.02}$&$8.91\pm0.20$&228.0(176)&$614$& xmm 5/(MOS1+MOS2)\\
  &                      &                      &             &          &  & +xmm 6/pn\\ 
\hline
28&$2.23^{+0.05}_{-0.07}$&$0.12\pm0.02$&$1.94\pm0.05$&81.7(91)&$134$& xmm 1/(MOS1+MOS2+pn)\\
  &$2.25^{+0.05}_{-0.06}$&$0.12\pm0.02$&$1.87\pm0.05$&132.7(118)&$129$& xmm 4/(MOS2+pn)\\
  &$2.13\pm0.06$&$0.11^{+0.02}_{-0.01}$&$1.62\pm0.04$&84.8(87)&$112$& xmm 6/pn\\
\hline
29&$2.12\pm0.04$&$0.14^{+0.02}_{-0.01}$&$2.80\pm0.07$&145.4(164)&$194$& xmm 1/(MOS1+MOS2+pn)\\
  &$1.82^{+0.08}_{-0.07}$&$0.10^{+0.03}_{-0.01}$&$4.52\pm0.20$&36.0(46)&$313$& xmm 2/pn\\
  &$1.61^{+0.03}_{-0.02}$&$0.09\pm0.01$&$6.50\pm0.11$&320.4(333)&$450$& xmm 4/(MOS1+MOS2+pn)\\
  &$1.96\pm0.04$&$0.16^{+0.01}_{-0.02}$&$4.19\pm0.09$&315.0(282)&$290$& xmm 6/(MOS1+MOS2+pn)\\
\hline
30&$2.87^{+0.45}_{-0.28}$&$0.15^{+0.07}_{-0.04}$&$0.35\pm0.03$&34.9(36)&$25$& xmm 4/pn\\
\hline
31&$1.56\pm0.02$&$0.09\pm0.01$&$15.95\pm0.18$&483.4(463)&$1103$& xmm 1/(MOS1+MOS2+pn)\\
  &$1.68^{+0.04}_{-0.03}$&$0.10\pm0.01$&$16.56\pm0.38$&163.0(202)&$1144$& xmm 2/(MOS1+MOS2+pn)\\
  &$1.71^{+0.01}_{-0.03}$&$0.10\pm0.01$&$13.31\pm0.18$&554.7(571)&$920$& xmm 4/(MOS1+MOS2+pn)\\
  &$1.66\pm0.02$&$0.09\pm0.01$&$12.53\pm0.16$&731.6(759)&$866$& xmm 6/(MOS1+MOS2+pn)\\
\hline
32&$0.67\pm0.04$         &$0.12\pm0.03$         &$18.64\pm0.38$&207.0(182)&$1288$& xmm 1/(MOS1+MOS2+pn)\\
  &$0.74^{+0.06}_{-0.05}$&$0.04^{+0.03}_{-0.01}$&$20.03\pm0.66$&  90.6(48)&$1384$& xmm 2/pn            \\
  &$0.56\pm0.02$         &$<0.05$               &$26.07\pm0.34$&412.4(348)&$1802$& xmm 4/(MOS1+MOS2+pn)\\
  &$0.57\pm0.03$         &$0.04\pm0.02$         &$10.82\pm0.19$&326.0(249)& $748$& xmm 6/(MOS1+MOS2+pn)\\
\hline
34&$2.08^{+0.32}_{-0.43}$&$0.10^{+0.05}_{-0.04}$&$0.27\pm0.02$&53.1(62)&$19$& xmm 1/pn+xmm 6/pn\\
\hline
36&$1.36^{+0.39}_{-0.26}$&$0.23^{+0.23}_{-0.11}$&$0.66\pm0.07$&20.7(33)&$46$& xmm 6/pn\\
\hline
37&$1.41^{+0.01}_{-0.05}$&$0.07\pm0.01$&$10.87\pm0.17$&181.8(200)&$751$& xmm 1/(MOS1+MOS2+pn)\\
  &$1.55^{+0.10}_{-0.09}$&$0.08\pm0.02$&$8.77\pm0.34$&33.8(39)&$606$& xmm 2/pn\\
  &$1.44^{+0.03}_{-0.02}$&$0.09\pm0.01$&$10.85\pm0.15$&317.1(282)&$750$& xmm 4/(MOS1+MOS2+pn)\\
  &$1.68\pm0.03$&$0.12^{+0.01}_{-0.02}$&$9.24\pm0.10$&401.7(352)&$640$& xmm 6/(MOS1+MOS2+pn)\\
\hline
38&$1.61^{+0.14}_{-0.13}$&$0.23\pm0.05$&$0.75\pm0.05$&33.2(32)&$52$&xmm 5/pn\\
\hline
39&$1.61^{+0.11}_{-0.04}$&$0.04\pm0.03$&$0.64\pm0.03$&26.3(26)&$44$&xmm 5/(MOS1+MOS2)\\
\hline
40&$1.90\pm0.32$&$0.19\pm0.07$&$0.54\pm0.05$&14.6(16)&$37$&xmm 6/pn\\
\hline
43&$1.56\pm0.04$&$0.13\pm0.01$&$21.80\pm0.30$&238.3(199)&$1507$&xmm 9/(MOS1+MOS2+pn)\\
\hline
\end{tabular}

\begin{list}{}{}
\item $^{a}$ -- Source number in Table \ref{source_ID}
\item $^{b}$ -- Equivalent hydrogen column depth in units of $10^{22}$ cm$^{-2}$ 
\item $^{c}$ -- Absorbed model flux in the $0.3 - 10$ keV energy range in 
units of $10^{-13}$ erg s$^{-1}$ cm$^{-2}$
\item $^{d}$ -- Absorbed isotropic source luminosity in the $0.3 - 10.0$ keV 
energy range in units of $10^{35}$ erg s$^{-1}$ assuming the distance of 760 
kpc
\end{list}
\end{table}

\clearpage

\begin{table}
\small
\caption{Best-fit model parameters of the energy spectra of the bright GC sources. 
{\em Chandra}/ACIS data, $0.5 - 7.0$ keV energy range. Absorbed simple power law 
model approximation. Parameter errors correspond to the $1 \sigma$ level. 
\label{spec_par_GCS_chandra_powerlaw}}
\begin{tabular}{ccccccc}
\hline
\hline
ID $^{a}$ & & & & & & Observation/Instrument \\
\hline
              & \multicolumn{5}{c}{Model: Absorbed Power Law (POWERLAW*WABS)}&        \\
\hline
              & Photon   & N$_{\rm H}^{b}$           &Flux$^{c}$&$\chi^{2}$&$L_{X}^{d}$&     \\
              & Index    &                           &          & (dof)    & &    \\
\hline
 6&$2.04^{+0.31}_{-0.17}$&$<0.08$               & $0.49\pm0.04$&  16.8(17)&  $34$& 2017/ACIS \\
\hline
 7&$1.93^{+0.44}_{-0.25}$&$0.60^{+0.33}_{-0.20}$& $0.73\pm0.06$&  31.2(23)&  $51$& 2017/ACIS \\
\hline
 8&$2.29^{+0.19}_{-0.16}$&$0.35^{+0.06}_{-0.05}$& $1.15\pm0.06$&  21.3(26)&  $80$& 1575/ACIS \\
\hline
 9&$2.07^{+0.08}_{-0.09}$&$0.51^{+0.08}_{-0.06}$& $2.07\pm0.08$&  20.9(39)& $143$& 1575/ACIS \\
\hline
10&$1.47\pm0.15$         &$0.56\pm0.14$         &$25.35\pm1.00$&  27.4(28)&$1752$& 1580/ACIS \\
  &$1.48^{+0.15}_{-0.14}$&$0.64^{+0.10}_{-0.14}$&$28.48\pm1.07$&  27.6(25)&$1968$& 1584/ACIS \\
  &$1.46^{+0.03}_{-0.04}$&$0.49\pm0.03$         &$33.72\pm0.40$&238.6(176)&$2330$& 2017/ACIS \\
  &$1.49^{+0.11}_{-0.13}$&$0.54^{+0.13}_{-0.11}$&$28.57\pm1.05$&  31.8(32)&$1975$& 2901/ACIS \\
\hline
11&$1.44\pm0.06$         &$0.14\pm0.03$         & $7.87\pm0.18$&  86.4(65)& $544$& 1575/ACIS \\
\hline
14&$2.40^{+0.30}_{-0.19}$&$0.31^{+0.09}_{-0.04}$& $0.91\pm0.05$&  19.6(15)&  $63$& 1575/ACIS \\
\hline
16&$1.73^{+0.11}_{-0.09}$&$0.24^{+0.04}_{-0.03}$& $3.94\pm0.14$&  37.2(34)& $272$& 1575/ACIS \\
\hline
20&$1.90^{+0.17}_{-0.30}$&$0.20^{-0.12}_{+0.03}$& $0.50\pm0.04$&  10.8(10)&  $35$& 1575/ACIS \\
\hline
22&$1.48\pm0.08$         &$0.22^{+0.03}_{-0.02}$&$49.48\pm1.23$&  60.8(62)&$3420$&  309/ACIS \\
  &$1.66^{+0.12}_{-0.11}$&$0.35^{+0.06}_{-0.10}$&$47.61\pm1.46$&  34.8(38)&$3290$& 1582/ACIS \\
  &$1.51\pm0.03$         &$0.30\pm0.02$         &$59.80\pm0.56$&261.7(215)&$4130$& 1575/ACIS \\
\hline
23&$1.71^{+0.17}_{-0.12}$&$<0.06$               & $0.54\pm0.04$&  14.6(19)&  $38$& 1575/ACIS \\
\hline
24&$1.84\pm0.05$         &$0.15^{+0.02}_{-0.01}$& $6.02\pm0.12$&  90.3(84)& $416$& 1575/ACIS \\
\hline
25&$1.36\pm0.05$         &$0.07\pm0.01$         & $7.87\pm0.15$&  78.8(84)& $544$& 1575/ACIS \\
\hline
26&$0.91^{+0.07}_{-0.06}$&$0.04^{+0.04}_{-0.03}$& $8.37\pm0.25$&  76.4(61)& $578$& 1575/ACIS \\
\hline
27&$2.02^{+0.07}_{-0.06}$&$0.04\pm0.02$         & $1.42\pm0.05$&  33.0(25)&  $98$& 1575/ACIS \\
\hline
29&$1.84^{+0.25}_{-0.10}$&$0.17^{+0.10}_{-0.07}$& $2.64\pm0.16$&  20.4(25)& $183$&  303/ACIS \\
  &$1.96\pm0.07$         &$0.13^{+0.02}_{-0.01}$& $3.65\pm0.09$&  76.6(56)& $252$& 1575/ACIS \\
\hline
30&$1.69^{+0.24}_{-0.27}$&$0.26^{+0.18}_{-0.09}$& $0.43\pm0.04$&  12.6(16)&  $30$& 1575/ACIS \\
\hline
31&$1.41^{+0.08}_{-0.04}$&$0.11\pm0.03$         &$15.88\pm0.44$&  54.7(59)&1097$$&  303/ACIS \\
  &$1.52\pm0.04$         &$0.06\pm0.01$         &$14.21\pm0.20$&210.4(171)& $982$& 1575/ACIS \\
\hline
32&$0.60^{+0.10}_{-0.11}$&$<0.08$               &$16.27\pm0.99$&  18.3(20)&$1125$&  310/ACIS \\
  &$0.26\pm0.14$         &$0.03^{+0.11}_{-0.03}$&$27.23\pm1.52$&  10.3(19)&$1880$& 1854/ACIS \\
  &$0.59^{+0.11}_{-0.09}$&$0.26^{+0.04}_{-0.16}$& $8.68\pm0.34$&  82.1(85)& $600$& 2017/ACIS \\
  &$0.50^{+0.14}_{-0.11}$&$0.18^{+0.13}_{-0.11}$&$14.12\pm0.67$&  47.2(33)& $976$& 2494/ACIS \\
\hline
35&$1.54^{+0.14}_{-0.13}$&$0.11\pm0.04$         &$15.30\pm0.71$&  30.9(23)&$1057$& 1582/ACIS \\
\hline
37&$1.40\pm0.04$         &$<0.04$               &$10.32\pm0.23$& 102.8(93)& $713$& 1575/ACIS \\
\hline
41&$1.44^{+0.15}_{-0.10}$&$<0.09$               &$16.02\pm0.69$&  29.3(28)&$1107$& 2895/ACIS \\
  &$1.55^{+0.19}_{-0.14}$&$0.04^{+0.06}_{-0.01}$&$16.84\pm0.88$&  10.9(17)&$1164$& 2896/ACIS \\
  &$1.58^{+0.16}_{-0.15}$&$0.08^{+0.07}_{-0.05}$&$16.35\pm0.77$&  22.2(22)&$1130$& 2897/ACIS \\
\hline
42&$1.66\pm0.06$         &$0.28\pm0.03$         &$74.50\pm1.43$&134.2(108)&$5148$& 2052/ACIS \\
  &$1.75^{+0.07}_{-0.06}$&$0.28^{+0.03}_{-0.02}$&$150.0\pm3.0$&102.9(69)&$10372$& 2053/ACIS \\
\hline
\end{tabular}

\begin{list}{}{}
\item $^{a}$ -- Source number in Table \ref{source_ID}
\item $^{b}$ -- Equivalent hydrogen column depth in units of $10^{22}$ cm$^{-2}$ 
\item $^{c}$ -- Absorbed model flux in the $0.3 - 10$ keV energy range in 
units of $10^{-13}$ erg s$^{-1}$ cm$^{-2}$
\item $^{d}$ -- Absorbed isotropic source luminosity in the $0.3 - 10.0$ keV 
energy range in units of $10^{35}$ erg s$^{-1}$ assuming the distance of 760 
kpc
\end{list}
\end{table}

\begin{table}
\small
\caption{Bright GCS spectral fit results ({\em XMM-Newton}/EPIC data, $0.3 - 10$ keV energy range; {\em Chandra}/ACIS data, 
$0.5 - 7.0$ keV energy range). Absorbed cut-off power law and Comptonization models. Parameter errors correspond to $1 \sigma$ 
level. \label{spec_par_GCS_cutoffpl_comptt}}
\begin{tabular}{ccccccccl}
\hline
\hline
ID $^{a}$ & & & & & & & & Remarks\\
\hline
              & \multicolumn{7}{c}{Model: Absorbed Cutoff Power Law (CUTOFFPL*WABS)} & \\
\hline
              & Photon & Cutoff Energy & &$N_{\rm H}^{b}$ & Flux$^{c}$ & $\chi^{2}$ & $L_{X}^{d}$& \\
              & Index  &   (keV)       & &                &            & (dof)      &            & \\
\hline
22&$0.64^{+0.13}_{-0.16}$&$3.64^{+0.80}_{-0.55}$&&$0.16^{+0.03}_{-0.02}$&$53.07\pm0.52$&228.2(215)&$3668$&{\em chandra 1575}\\
  &$0.83^{+0.13}_{-0.10}$&$4.13^{+0.84}_{-0.50}$&&$0.16^{+0.02}_{-0.01}$&$47.74\pm0.40$&399.3(423)&$3300$&{\em xmm 5}\\
\hline
26&$0.18\pm0.09$         &$4.86^{+1.12}_{-0.72}$&&$<0.02$               & $7.91\pm0.17$&208.0(175)& $545$&{\em xmm 5+xmm 6}\\
\hline
32&$-0.15\pm0.09$        &$4.40^{+0.46}_{-0.53}$&&$<0.03$               &$16.71\pm0.28$&171.8(181)&$1155$&{\em xmm 1}\\
  &$0.17\pm0.15$         &$5.43^{+2.05}_{-1.20}$&&$<0.02$               &$18.59\pm0.61$&  77.4(47)&$1285$&{\em xmm 2}\\
  &$0.21\pm0.07$         &$7.91^{+1.75}_{-1.15}$&&$<0.01$               &$23.71\pm0.31$&376.6(347)&$1640$&{\em xmm 4}\\
  &$0.01^{+0.05}_{-0.08}$&$5.74^{+0.65}_{-1.05}$&&$<0.01$               &$10.69\pm0.17$&278.0(248)& $739$&{\em xmm 6}\\
\hline
42&$0.83^{+0.32}_{-0.24}$&$3.83^{+2.35}_{-0.97}$&&$0.16\pm0.05$         &$68.39\pm1.32$&126.4(107)&$4726$&{\em chandra 2052}\\
  &$0.13^{+0.31}_{-0.32}$&$1.81^{+0.45}_{-0.31}$&&$0.09\pm0.03$         & $125.9\pm2.6$&  74.2(68)&$8700$&{\em chandra 2053}\\
\hline
43&$1.13^{+0.17}_{-0.24}$&$6.82^{+4.47}_{-1.94}$&&$0.09\pm0.02$         &$20.39\pm0.29$&232.0(198)&$1409$&{\em xmm 9}\\
\hline
              & \multicolumn{7}{c}{Model: Absorbed Comptonization Model (COMPTT*WABS)} & \\
\hline
              & $kT_{0}$ & $kT_{e}$ & $\tau$        & $N_{\rm H}^{b}$& Flux$^{c}$ & $\chi^{2}$&$L_{X}^{d}$& \\
              &  (keV)   &  (keV)   &               &                &   & (dof) & & \\   
\hline
22&$0.20^{+0.04}_{-0.15}$&$1.46^{+0.08}_{-0.07}$&$26.3^{+1.6}_{-2.0}$&$0.17^{+0.09}_{-0.07}$&$50.29\pm0.49$&229.2(217)&$3475$&{\em chandra 1575}\\
  &$0.37^{+0.01}_{-0.03}$&$1.88^{+0.23}_{-0.18}$&$19.5^{+0.9}_{-1.6}$&$<0.09$&$47.82\pm0.41$&398.7(422)&$3297$& {\em xmm 5}\\
\hline
26&$0.03^{+0.07}_{-0.01}$&$1.76^{+0.14}_{-0.12}$&$34.6^{+2.8}_{-2.5}$&$0.07^{+0.01}_{-0.02}$&$7.38\pm0.16$&206.8(174)&$509$& {\em xmm 5+xmm 6}\\
\hline
32&$0.06^{+0.05}_{-0.06}$&$1.87^{+0.12}_{-0.10}$&$38.6^{+2.9}_{-2.7}$&$0.10^{+0.03}_{-0.04}$&$16.11\pm0.27$&177.0(180)&$1113$& {\em xmm 1}\\
  &$0.07^{+0.06}_{-0.06}$&$1.63^{+0.13}_{-0.10}$&$44.8^{+3.8}_{-4.6}$&$<0.09$               &$17.63\pm0.58$&  65.5(46)&$1220$& {\em xmm 2}\\
  &$0.09^{+0.02}_{-0.08}$&$1.82^{+0.08}_{-0.06}$&$43.2^{+1.3}_{-2.1}$&$<0.06$               &$21.63\pm0.28$&341.7(346)&$1495$& {\em xmm 4}\\ 
  &$0.10^{+0.05}_{-0.09}$&$1.79^{+0.08}_{-0.07}$&$45.2^{+3.7}_{-2.7}$&$0.03^{+0.03}_{-0.01}$& $9.93\pm0.16$&252.3(247)& $686$& {\em xmm 6}\\
\hline
42&$0.17^{+0.05}_{-0.06}$&$1.45^{+0.25}_{-0.14}$&$23.7^{+2.9}_{-1.7}$&$0.19^{+0.03}_{-0.06}$&$65.46\pm1.26$&125.7(106)&$4524$&{\em chandra 2052}\\
  &$<0.14$&$1.08^{+0.10}_{-0.08}$&$29.5^{+3.5}_{-3.0}$&$0.18\pm0.03$&$122.8\pm2.5$&71.2(67)&$8485$&{\em chandra 2053}\\
\hline
43&$0.14\pm0.04$&$1.64^{+0.15}_{-0.16}$&$22.4^{+2.2}_{-2.3}$&$0.09\pm0.02$&$19.42\pm0.28$&258.3(235)&$1339$& {\em xmm 9}\\
\hline 
\end{tabular}

\begin{list}{}{}
\item $^{a}$ -- Source number in Table \ref{source_ID}
\item $^{b}$ -- Equivalent hydrogen column depth in units of $10^{22}$ cm$^{-2}$ 
\item $^{c}$ -- Absorbed model flux in the $0.3 - 10$ keV energy range in 
units of $10^{-13}$ erg s$^{-1}$ cm$^{-2}$
\item $^{d}$ -- Absorbed isotropic source luminosity in the $0.3 - 10.0$ keV 
energy range in units of $10^{35}$ erg s$^{-1}$ assuming the distance of 760 
kpc
\end{list}
\end{table}

\begin{table}
\small
\caption{Bright GCS spectral fit results, {\em XMM-Newton}/EPIC and {\em Chandra}/ACIS data. Two-component model 
approximation: (BBODYRAD+DISKBB)*WABS and (BBODYRAD+POWERLAW)*WABS models. Parameter errors correspond to $1 \sigma$ 
level.  \label{spec_par_GCS_two_comp}}
\begin{tabular}{ccccccccl}
\hline
\hline
ID $^{a}$ & & & & & & & & Observation\\
\hline
       & \multicolumn{7}{c}{Model: (BBODYRAD+DISKBB)*WABS} & \\
\hline
       & $kT_{BB}$ & $kT_{in}$&$r_{in} \sqrt{cos i}$    &$N_{\rm H}^{b}$& Flux$^{c}$ & $\chi^{2}$&$L_{X}^{d}$& \\
       &  (keV)    &  (keV)   &      (km)               &               &            &   (dof)   &           & \\   
\hline
22&$1.40^{+0.25}_{-0.16}$&$0.83\pm0.16$&$37^{+13}_{-8}$&$0.17\pm0.03$&$50.94\pm0.50$&229.3(217)&$3520$&{\em chandra 1575}\\
  &$1.84^{+0.25}_{-0.27}$&$1.05^{+0.16}_{-0.14}$&$26^{+5}_{-4}$&$0.14\pm0.01$&$48.26\pm0.42$&395.1(421)&$3335$&{\em xmm 5}\\
\hline
42&$...$&$1.70\pm0.08$&$16^{+1}_{-2}$&$0.10\pm0.02$&$65.31\pm1.25$&128.2(106)&$4513$&{\em chandra 2052}\\
  &$...$&$1.45\pm0.07$&$30\pm3$&$0.12\pm0.02$&$125.4\pm2.50$&71.2(67)&$8665$&{\em chandra 2053}\\
\hline
43&$1.30^{+0.12}_{-0.11}$&$0.55\pm0.02$&$45^{+11}_{-8}$&$0.09\pm0.02$&$19.48\pm0.27$&258.0(233)&$1346$&{\em xmm 9}\\
\hline
       & \multicolumn{7}{c}{Model: (BBODYRAD+POWERLAW)*WABS} & \\
\hline
       & $kT_{BB}$ &Photon &$r_{BB}$    &$N_{\rm H}^{b}$& Flux$^{c}$ & $\chi^{2}$&$L_{X}^{d}$& \\
       &  (keV)    &Index  &   (km)     &               &            &   (dof)   &           & \\   
\hline
22&$1.10^{+0.09}_{-0.12}$&$1.82^{+0.28}_{-0.22}$&$27\pm3$      &$0.28\pm0.05$&$52.45\pm0.52$&229.0(217)&$3625$&{\em chandra 1575}\\
  &$0.82^{+0.07}_{-0.05}$&$1.56^{+0.07}_{-0.06}$&$28^{+6}_{-4}$&$0.21^{+0.01}_{-0.02}$&$49.87\pm0.43$&392.6(421)&$3446$&{\em xmm 5}\\
\hline
42&$0.87^{+0.26}_{-0.13}$&$1.67^{+0.44}_{-0.12}$&$23^{+6}_{-8}$  &$0.22^{+0.06}_{-0.05}$&$70.59\pm1.35$&125.2(106)&$4878$&{\em chandra 2052}\\
  &$0.91^{+0.05}_{-0.04}$&$2.66^{+0.77}_{-0.43}$&$82^{+12}_{-13}$&$0.32^{+0.11}_{-0.06}$&$122.9\pm2.5$ &71.3(67)  &$8493$&{\em chandra 2053}\\
\hline
43&$1.20^{+0.13}_{-0.12}$&$1.91^{+0.33}_{-0.19}$&$16^{+2}_{-3}$&$0.16^{+0.04}_{-0.02}$&$20.20\pm0.28$&215.2(195)&$1396$&{\em xmm 9}\\
\hline
\end{tabular}

\begin{list}{}{}
\item $^{a}$ -- Source number in Table \ref{source_ID}
\item $^{b}$ -- Equivalent hydrogen column depth in units of $10^{22}$ cm$^{-2}$ 
\item $^{c}$ -- Absorbed model flux in the $0.3 - 10$ keV energy range in 
units of $10^{-13}$ erg s$^{-1}$ cm$^{-2}$
\item $^{d}$ -- Absorbed isotropic source luminosity in the $0.3 - 10.0$ keV 
energy range in units of $10^{35}$ erg s$^{-1}$ assuming the distance of 760 
kpc
\end{list}
\end{table}

\clearpage

\clearpage

\clearpage

\clearpage

\clearpage

\clearpage

\clearpage

\clearpage

\clearpage

\clearpage

\clearpage

\clearpage

\clearpage

\clearpage

\clearpage

\clearpage

\clearpage

\clearpage

\end{document}